\documentclass[aps,amsmath,amssymb,preprintnumbers,preprint,nofootinbib,a4paper,11pt]{revtex4-1}
\pdfoutput=1
\usepackage[utf8]{inputenc} 
\usepackage{graphicx}
\usepackage{url}
\usepackage[bookmarks, pagebackref=false]{hyperref}
\usepackage[usenames,dvipsnames]{xcolor}
\definecolor{orange}{cmyk}{0,0.5,1,0}
\definecolor{rossoCP3}{cmyk}{0,.88,.77,.40}
\definecolor{graa}{rgb}{0.8,0.8,0.8}
\definecolor{blaa}{rgb}{0.2,0.2,0.6}
\hypersetup{
	colorlinks, 
	bookmarksopen, 
	bookmarksnumbered,
	citecolor=blaa, 		
	linkcolor=rossoCP3,	
	urlcolor=rossoCP3,			
	    }

\usepackage{amsthm}
\usepackage{bm}
\usepackage{bbm}
\usepackage{pxfonts}

\usepackage{amsmath,amssymb,amsfonts}
\usepackage{color}
\usepackage{float}
\usepackage{hyperref}
\usepackage[Symbolsmallscale]{upgreek}
\usepackage{amsmath}
\usepackage{amsfonts}
\usepackage{amssymb,dsfont}
\usepackage{graphicx}
\usepackage{amssymb}
\usepackage[vcentermath]{youngtab}
\usepackage[all]{xy}
\usepackage{pstricks}
\usepackage{dsfont}%
\usepackage{multirow}

\usepackage{cases}
\usepackage{comment}
\usepackage{enumerate}

\usepackage{placeins}
\usepackage{xspace}
\usepackage{cancel} 

\usepackage{slashed}
\usepackage[caption=false, labelformat=simple, listofformat=subsimple, labelfont=default, margin=5pt,
    justification=raggedright]{subfig}

\makeatletter
	\renewcommand{\p@subfigure}{}
\makeatother

\usepackage{natbib}

\usepackage{feynmf}

\newcommand{\newc}{\newcommand}
\newc{\renewc}{\renewcommand}

\newcommand{\su}[1]{\text{SU(#1)}}
\newcommand{\uu}[0]{\text{U(1)}}
\newcommand{\tred}[1]{{\color{red} #1}}
\newcommand{\tblu}[1]{{\color{blue} \bf #1}}
\newcommand{\even}[0]{{\color{blue} even}}
\newcommand{\odd}[0]{{\color{red} odd}}

\begin{document}
\thispagestyle{empty}

\title{\Large Systematic classification of aGUT models in five dimensions: \\ The SU(N) kinship}

\author{Giacomo Cacciapaglia}
\email{g.cacciapaglia@ip2i.in2p3.fr}
\affiliation{Institut de Physique des 2 Infinis de Lyon (IP2I), UMR5822,
	CNRS/IN2P3, F-69622 Villeurbanne Cedex, France}
\affiliation{University of Lyon, Universit\'e Claude Bernard Lyon 1, F-69001 Lyon, France}

\begin{abstract}

Asymptotic Grand Unification theories (aGUTs) in five dimensions provide a valid alternative to standard quantitative unification. We define the pathway towards viable models starting from a general unified bulk gauge symmetry. Imposing the presence of ultra-violet fixed points for both gauge and Yukawa couplings strongly limits the possibilities. Within the SU(N) kinship, we identify and characterise only two realistic minimal models, both based on a bulk SU(6) symmetry. Both models feature the generation of either up or down-type Yukawas via gauge scalars, two Higgs doublets with build-in minimal flavour violation at low energies, and conservation of baryon number. We also propose interesting avenues beyond the minimality criterion.

\end{abstract}

\maketitle

\section{Introduction}


In spite of the precision tests of the Standard Model (SM) of particle physics and the recent discovery of its last accounted particle, the Higgs boson \cite{ATLAS:2012yve,CMS:2012qbp}, the theory falls short of the status of ultimate fundamental theory in three directions. Firstly, some experimental data cannot be explained, \emph{in primis} the presence of dark matter  \cite{Bertone:2018krk} (and dark energy) in today's Universe. Secondly, the SM lacks an explanation of the dynamical origin of the Higgs mechanism, gravitational interactions and a dynamical connection between the electroweak and gravitational scales. Thirdly, while based on simple principles, the final construction lacks elegance in both the gauge sector and the three-family structure of the fermions. In fact, the three gauge factors of the SM have very different dynamical properties: while $\su 3_c$ is asymptotically free \cite{Politzer:1973fx,Gross:1973id} and dynamically confines at energies below the GeV, the $\su 2_L$ is broken by the vacuum \cite{Englert:1964et,Higgs:1964pj}, while the $\uu_Y$ runs into a Landau pole in the Ultra-Violet (UV). A simple and elegant solution would be to assume that the SM gauge structure is replaced by a single simple gauge group at higher energy, which encompasses the whole SM gauge structure and leads to Grand Unification of gauge forces \cite{Georgi:1974sy}. 
The traditional approach to building Grand Unified Theories (GUTs) originates from the observation that the gauge couplings tend to similar values at high energies under the renormalisation group evolution. Hence, quantitative unification is expected at a given scale $\Lambda_\text{GUT}$, above which an extended gauge symmetry is recovered, for instance $\su 5$ \cite{Georgi:1974sy} or SO(10) \cite{FRITZSCH1975193}. At the scale $\Lambda_\text{GUT}$, the grand unified gauge symmetry needs to be broken via an analogue of the Higgs mechanism in the SM, often requiring many scalar fields in large representations \cite{Aulakh:2003kg}.

In this work, we follow an alternative approach inspired by the recent developments in asymptotic safety in quantum field theory \cite{Litim:2014uca,Litim:2015iea}: instead of meeting at a fixed scale, the running gauge couplings tend to the same fixed point in the UV \cite{Bajc:2016efj}.~\footnote{Models with both supersymmetry \cite{Bajc:2016efj,Bajc:2023uls} and without \cite{Molinaro:2018kjz,Fabbrichesi:2020svm} have been constructed.} This new approach leads to asymptotic Grand Unification Theories (aGUTs) \cite{Cacciapaglia:2020qky}. A simple way to realise aGUTs consists on building theories in five dimensions (5D). The common lore is that 5D theories have a natural cut-off due to the linear running of the gauge couplings \cite{Contino:2001si}, which renders them intrinsically non-renormalisable. However, under certain conditions, the gauge running in the UV is tamed by the presence of a fixed point, which renders the theory renormalisable \cite{Gies:2003ic,Morris:2004mg} and, therefore, valid up to arbitrary scales. For this, it suffices that the one-loop beta function is negative, i.e. it would lead to an asymptotically-free theory in 4 dimensions (4D). Contrary to the traditional unification, in aGUTs an exact unified gauge invariance is absent at all scales, but it is approached with increasing precision at high energies, asymptotically. What the model provides is a scale, $\Lambda_\text{aGUT} \equiv m_\text{KK}$, where the running of the gauge couplings is modified and which coincides with the mass of the first Kaluza-Klein (KK) resonance \cite{Cacciapaglia:2020qky}. As the fixed point dynamics occurs at scales much higher that the KK scale, its presence does not depend on the detailed spectrum of the low lying states, hence it occurs both in flat and warped 5D models \cite{Randall:1999ee}.

While a UV fixed point is relatively easy to obtain in the gauge sector, Yukawa couplings are more problematic as their running can easily incur into a Landau pole right above the KK scale. The first aGUT models have been constructed based on a bulk $\su 5$ \cite{Cacciapaglia:2020qky,Cacciapaglia:2022nwt} and SO(10) \cite{Khojali:2022gcq} symmetry. For the former, the dynamics of the Yukawa couplings poses strong limitations in the model building, pointing towards non-minimal configurations, while the SO(10) case has been proven to be non-viable. In this perspective, gauge-Higgs unification mechanism \cite{Hosotani:1983xw,Contino:2003ve} comes in handy, as it allows to link a Yukawa coupling to the bulk gauge couplings \cite{Hosotani:2015hoa,Nomura:2006pn}. This stems from the identification of the Higgs boson with a scalar component of the 5D gauge field (gauge-scalar). Hence, the Yukawa coupling will inherit the same attractive fixed point as the gauge ones, being identical at short distances. Supersymmetry can also link Yukawa couplings to the bulk gauge coupling if some of the SM fermions emerge from the bulk gauginos, as shown in the case of $\text{E}_6$ aGUT \cite{Cacciapaglia:2023ghp}. All-in-all, aGUT model building offers several avenues, however the strong constraints from the UV behaviour of the theory tightly limits the number of possibilities.

In this work we aim at defining the pathway towards a systematic classification of aGUT models. As we will show, the tight requirements aGUTs need to satisfy lead to a handful of viable models, in particular limiting the rank of the bulk gauge symmetry. Hence, it is possible to compile a complete and exhaustive catalogue of aGUTs, whose dynamics and low-energy predictions can be studied systematically. We will apply the pathway to models with an SU(N) symmetry in the bulk, showing that only two viable models are possible. As we will see, the UV behaviour of the bulk Yukawa couplings and the introduction of three generations play a crucial role in shaping the aGUT models.

After introducing the rules of the game in Section~\ref{sec:pathway}, we show how to search for aGUTs within the SU(N) kinship, ruled by the classic $\su 5$, in Section~\ref{sec:sun}. Finally, Section~\ref{sec:Yukawa} is dedicated to a detailed study of the Yukawa sector in the viable models, ruling out all but two. Finally, in Section~\ref{sec:concl} we offer our conclusions and discuss the perspectives for the completion of the aGUT catalogue.

\section{The minimal pathway} \label{sec:pathway}

The aGUT models are based on a 5D background, where one extra spatial dimension is compactified on an orbifold. Here, by orbifold we mean a compact space obtained by the action of a discrete symmetry group, hence for a single spatial dimension the maximal group consists in two parities, $\mathbb{Z}_2 \times \mathbb{Z}_2'$.  The bulk gauge group is then broken by an inner automorphism \cite{Hebecker:2003jt}, i.e. the action of the discrete symmetry group combined with the action of the bulk gauge group on itself. This is the only way to preserve the bulk gauge symmetry at high scales, see Appendix~\ref{app:BCs} for a more detailed discussion. Hence, the model is defined in terms of the bulk gauge group SU(N) and two parities, together with the parity assignments for all bulk fields. As already mentioned, the curvature of the bulk is not relevant for the UV behaviour, hence the aGUT model building can proceed equally in flat or warped space.

To classify all possible aGUTs, we establish the following procedure:
\begin{itemize}
    \item[A)] Define a bulk gauge symmetry $\mathcal{G} \supset \mathcal{G}_\text{SM} = \su 3_c \times \su 2_L \times \uu_Y$.

    \item[B)] Identify the parities $P_i$, $i=0,\dots n$, that allow to break $\mathcal{G} \to \mathcal{H}_i \supset \mathcal{G}_{\rm SM}$. Note that $P_0$ labels the identity, which does not break the bulk gauge group ($\mathcal{H}_0 \equiv \mathcal{G}$).

    \item[C)] Define viable pairs of parities, $P_i\times P_j$, such that the remnant 4D gauge symmetry
    \begin{equation}
    \mathcal{G}_\text{4D} \equiv \mathcal{H}_i \cap \mathcal{H}_j = \mathcal{G}_\text{SM} \times X\,,
    \end{equation}
    where $X$ indicates an extra gauge group, which needs to be broken by a Higgs mechanism in the bulk or on the boundaries. The minimal requirement is $X = \uu^k$, while non-abelian gauge groups may be harder to efficiently break to nil.

    \item[D)] Find minimal sets of bulk fermions in irreducible representations (\emph{irreps}) of $\mathcal{G}$ that preserve the UV fixed point. Upon the application of the orbifold parities, the zero modes should consist of complete generations of SM fermions. Additional zero mode fermions are allowed if vector-like with respect to the SM gauge symmetry, provided they can acquire mass via interactions localised on the boundaries. 

    \item[E)] Write down all allowed Yukawa couplings with a bulk scalar containing the Higgs doublet as zero mode. Additional Yukawas may emerge from gauge-scalars, depending on the parities.

    \item[F)] Check the UV running of the Yukawa couplings: all of them must run to a UV fixed point.

    \item[G)] Complement the model with supersymmetry if some fermions can stem from the adjoint irrep. Check again for the presence of the gauge UV fixed point.

\end{itemize}

As we will discuss in Section~\ref{sec:Yukawa}, the UV fate of Yukawa couplings is crucial for the success of the model, and it also determines how the three generations can be introduced.

The minimalist nature of this pathway relies on two points: the remnant 4D gauge symmetry $\mathcal{H}_i \cap \mathcal{H}_j$ consists of the SM gauge symmetry; the bulk fermions only contain anomaly-free sets of SM fermions (i.e. complete families), so that no chiral states need to be added on the boundaries. The latter provides a huge simplification for the model building, as it is hard to provide mass for all unwanted bulk chiral fermions while respecting the gauge symmetry in the bulk and boundaries. The former, instead, could be extended by allowing some king of intermediate step between the SM gauge symmetry and the unified invariance, like for instance Pati-Salam \cite{Pati:1974yy}. We will discuss some possible extension in the concluding section.

Before plunging in the hunt of SU(N) aGUTs, we discuss in some more details the action of the orbifold and how to ensure the presence of a gauge UV fixed point. We will only provide the basic information needed for the follow-up exploration.

\subsection{Symmetry breaking by orbifold projection}

For each parity $P_i$, the action on the fields can be characterised in terms of a matrix, which is a specific transformation under the bulk gauge group $\mathcal{G}$. We indicate this matrix as $\mathcal{P}_i$, with the property that $\mathcal{P}_i \cdot \mathcal{P}_i = \mathcal{P}_0$. 

For the gauge fields, divided in the 4D polarisations $A_\mu$ and the fifth component $A_5$, the action of the parity reads:
\begin{equation} \label{eq:parityA}
    A_\mu \to \mathcal{P}_i \cdot A_\mu \cdot \mathcal{P}_i\,, \quad A_5 \to - \mathcal{P}_i \cdot A_5 \cdot \mathcal{P}_i\,.
\end{equation}
Hence the parity of the gauge-scalar components are opposite to those of the corresponding vector components. This is purely due to the 5-vector nature of the gauge field. Equation~\eqref{eq:parityA} provides a parity assignment, $\pm$, for each component of the fields under the 4D remnant $\mathcal{G}_\text{4D}$. In the following, we will explicitly report the parities of the 4-vector components, so that $(+,+)$ yields a 4D vector zero mode, while $(-,-)$ components feature a gauge-scalar zero mode.

For other bulk fields, the parity matrix $\mathcal{P}_i$ acts on the fundamental and anti-fundamental indices of the $\mathcal{G}$ irrep, providing an intrinsic parity assigned to each of the components under $\mathcal{G}_\text{4D}$. One can also define an overall field-specific parity $\eta = \pm 1$. For a generic scalar bulk field, the action of each parity reads:
\begin{equation}
    \Phi \to \eta_\phi \ (\mathcal{P}_{i}\ \dots \mathcal{P}_{i}) \cdot \Phi \cdot (\mathcal{P}_{i}\ \dots \mathcal{P}_{i})\,,
\end{equation}
where the $\mathcal{P}_i$ matrices on the left act on the $n$ fundamental indices while those on the right on the $m$ anti-fundamental indices of the irrep. Note that the parities provided by the matrices $\mathcal{P}_i$ are the same for all fields in the same irrep, while the overall sign $\eta_\phi = \pm 1$ is field specific. Note also that the gauge indices in the above equation are left understood and scalar zero modes are only present for components with $(+,+)$ parities.

Fermion field in 5D are intrinsically Dirac, i.e. 4-component spinors, hence they contain both left-handed (lh) and right-handed (rh) 4D polarisations. As the action of the parity distinguishes the 4D chirality, it is convenient to split the 5D spinor in its 4D chiral components:
\begin{equation}
    \Psi = \Psi_\text{lh} + \Psi_\text{rh}\,,
\end{equation}
where $\Psi_\text{lh} = \frac{1-\gamma^5}{2} \Psi$ and $\Psi_\text{rh} = \frac{1+\gamma^5}{2} \Psi$.
The action of the parity $P_i$ reads:
\begin{eqnarray}
    \Psi_\text{lh} &\to& \eta_\Psi\ (\mathcal{P}_i\ \dots \mathcal{P}_i) \cdot \Psi_\text{lh} \cdot (\mathcal{P}_i\ \dots \mathcal{P}_i)\,, \\
    \Psi_\text{rh} &\to& -\eta_\Psi\ (\mathcal{P}_i\ \dots \mathcal{P}_i) \cdot \Psi_\text{rh} \cdot (\mathcal{P}_i\ \dots \mathcal{P}_i)\,,
\end{eqnarray}
where we define an overall field-specific sign $\eta_\Psi = \pm 1$, and the parities of the two chiralities are opposite.
In the following, we will always indicate the parities for the left-handed components. Hence, $(+,+)$ yields a left-handed zero mode, while $(-,-)$ yields a right-handed zero mode, while components with $(+,-)$ and $(-,+)$ parities have no zero modes.

\subsection{Preserving the gauge UV fixed point} \label{subsec:fixed}

The existence of a UV fixed point depends on the power-law running of the 5D gauge coupling \cite{Dienes:2002bg,Hebecker:2004xx}. As it carries dimensions in mass, we define an effective 't Hooft coupling, which contains linearly the renormalisation scale $\mu$:
\begin{equation} \label{eq:alphatildedef}
\tilde{\alpha} = \alpha \mu R\,,
\end{equation}
where $R \equiv m_\text{KK}^{-1}$ is the radius of the extra dimension and $\alpha_i$ is the 4D effective coupling, $\alpha = g^2/4\pi$.
The 5D beta function at one-loop reads \cite{Cacciapaglia:2020qky}:
\begin{equation} \label{eq:5Derg}
2 \pi \frac{d}{d \ln \mu} \tilde{\alpha} = 2 \pi \tilde{\alpha} - b_5\ \tilde{\alpha}^2\,.
\end{equation}
For a non-supersymmetric bulk theory, we find
\begin{equation} \label{eq:b5}
b_5 = \left( \frac{11}{3} - \frac{1}{6}\right) C(G) - \frac{4}{3} \sum_f T(R_f) - \frac{1}{3} \sum_s T(R_s)\,,
\end{equation}
where $f$ and $s$ indicate the fermion and (complex) scalar bulk fields, irrespective of their boundary conditions (parity assignments). This effect is dominated by the contribution of the high-mass KK modes, which tend to reconstruct complete irreps of the bulk gauge symmetry \cite{Dienes:1996du}. Hence, to compute Eq.~\eqref{eq:b5} is if enough to consider the bulk group $\mathcal{G}$ and the complete bulk irreps.

The UV fixed point exists as long as $b_5>0$ and it is given by:
\begin{equation}
\tilde{\alpha}^\ast_\text{UV} = \frac{2 \pi}{b_5}\,.
\end{equation}
As the computation is perturbative, is it necessary that $\tilde{\alpha}^\ast_\text{UV}$ also remains perturbative \cite{Cacciapaglia:2020qky}.

Supersymmetric models in 5D effectively consist of $\mathcal{N}=2$ SUSY. Hence, the beta function can be computed by adding the necessary SUSY partners to Eq.~\eqref{eq:b5}. The one-loop beta function has also been computed directly in 5D leading to \cite{Flacke:2003ac}:
\begin{equation} \label{eq:b5SUSY}
b_5^{\rm SUSY} = \frac{\pi}{2} \left( C(G) - \sum_m T(R_m) \right)\,,
\end{equation}
where $m$ labels the matter super-multiplets. Note that the result in Eq.~\eqref{eq:b5SUSY} differs from Eq.~\eqref{eq:b5} only by the pre-factor \cite{Cacciapaglia:2023ghp}, hence the conditions on the bulk irreps remain independent on how the beta function is computed.

\section{Finding the SU(N) aGUTs} \label{sec:sun}

As described in Sec.~\ref{subsec:fixed}, the irreps one can use to build aGUT models are limited by the presence of the UV fixed point for the gauge coupling. In Table~\ref{tab:irreps}, we list all the irreps valid for SU(N) models. Except for the two-index irreps, higher dimensionality ones are only allowed for a limited number of colours N, as indicated in the last two columns. Note also that, following Eq.~\eqref{eq:b5SUSY}, SUSY models are much more restrictive, allowing only fundamentals, and two-index symmetric and anti-symmetric irreps, with a limited multiplicity.

\begin{table}[h!]
\centering
\begin{tabular}{|c|c|c|c|c|c|c|}
\hline
irrep & dimension & $T(R)$ & \multicolumn{2}{c|}{5D} & \multicolumn{2}{c|}{SUSY} \\
& & & $N_{max}$ & mult. & $N_{max}$ & mult. \\ \hline
Adj & $N^2-1$ & $N$ &  none & $2$ & -- & -- \\ \hline
F & $N$ & $1/2$ &  none & $\frac{21 N}{4}$ & none & $2N$ \\ \hline
A & $N (N-1)/2$ & $(N-2)/2$ &  none & $\frac{21 N}{4 (N-2)}$ & none & $\frac{2N}{N-2}$ \\ \hline 
S & $N (N+1)/2$ & $(N+2)/2$ &  none & $\frac{21 N}{4 (N+2)}$ & none & $\frac{2N}{N+2}$ \\ \hline 
A$_3$ & $N (N-1) (N-2)/6$ & $(N-2)(N-3)/4$ &  $15$ & $\frac{21 N}{4 (N-2)}$ & -- & -- \\ \hline 
A$_4$ & $N (N-1) (N-2) (N-3)/24$ & $(N-2)(N-3) (N-4)/12$ &  $8,\ 9$ & $2,\ 1$ & -- & -- \\ \hline
${\bf 40}$ & $N(N^2-1)/3$ & $(N^2-3)/2$ &  $5$ & 1 & -- & -- \\ \hline
${\bf 45}$ & $N (N+1)(N-2)/2$ & $(N-2)(3N+1)/4$ & $5$ & 1 & -- & -- \\ \hline
\end{tabular}
\caption{\label{tab:irreps} Irreps of SU(N) allowed by the UV fixed point in 5D and in the SUSY version. In the last two columns, $N_{max}$ indicates the largest number of colours for which this irrep is allowed, while the maximal multiplicity is indicated in the next column. Besides the adjoint (Adj), F indicates the fundamental, A the 2-index anti-symmetric, S the two index symmetric, A$_3$ and A$_4$ the three and four index anti-symmetric. The last two rows contain irreps that are only allowed for $\su 5$.}
\end{table}

The most general parity matrix $\mathcal{P}_i$ can be written as a diagonal with $\pm 1$ entries, hence breaking $\su{\text{N}} \to \su m \times \su{N-m} \times \uu$, with $m = 1, \dots [N/2]$, where $[x]$ indicates the integer part of $x$. We can now investigate individual SU(N) group, starting from the smallest rank $\su 5$.

\subsection{SU(5) (rank 4)} \label{subsec:su5}

$\su 5$ is the smallest rank SU(N) group that can accommodate the SM gauge symmetry, having rank 4 \cite{Georgi:1974sy}.
There is only one non-trivial parity that allows $\mathcal{H}_1 \supset \mathcal{G}_\text{SM}$:
\begin{equation}
    P_1: \quad \su 5 \to \su 3 \times \su 2 \times \uu \equiv \mathcal{G}_{\rm SM}\,.
\end{equation}
It is defined in terms of the matrix:
\begin{equation}
    \mathcal{P}_1 = \text{diag} (+,\ +,\ +,\ -,\ -)\,, \quad \mathcal{H}_1 = G_\text{SM}\,.
\end{equation}
The generator of the $\uu$, identified with hypercharge, is normalised as follows:
\begin{equation}
    Y =  \text{diag} (-\frac{1}{3},\ -\frac{1}{3},\ -\frac{1}{3},\ \frac{1}{2},\ \frac{1}{2})\,.
\end{equation}

The SU(5) irreps from Table~\ref{tab:irreps} decompose under $P_1$ as:
\begin{eqnarray}
    \text{Adj} = {\bf 24} &\to & \tblu{(8,1)_0} \oplus  \tblu{(1,3)_0} \oplus \tblu{(1,1)_0} \oplus \tred{(3,2)_{-5/6}} \oplus \tred{(\bar{3},2)_{5/6}}\,, \\
    \text{F} = {\bf 5} &\to & \tblu{(3,1)_{-1/3}} \oplus \tred{(1,2)_{1/2}} \,, \\
    \text{A} = {\bf 10} &\to & \tred{(3,2)_{1/6}} \oplus \tblu{(\bar{3},1)_{-2/3}} \oplus \tblu{(1,1)_{1}}\,, \\
    \text{S} = {\bf 15} &\to & \tred{(3,2)_{1/6}} \oplus \tblu{(6,1)_{-2/3}} \oplus \tblu{(1,3)_{1}}\,, \\
    {\bf 40} &\to & \tblu{(8,1)_{-1}} \oplus \tred{(\bar{3},2)_{-1/6}} \oplus \tred{(6,2)_{-1/6}} \oplus \tblu{(3,3)_{2/3}} \oplus \tblu{(3,1)_{2/3}} \oplus \tred{(1,2)_{3/2}}\,, \\
    {\bf 45} &\to & \tblu{(3,1)_{-4/3}} \oplus \tred{(1,2)_{-1/2}} \oplus \tred{(8,2)_{-1/2}}\oplus \tblu{(6,1)_{1/3}} \oplus \tblu{(\bar{3},3)_{1/3}} \oplus \tred{(3,2)_{7/6}} \oplus \tblu{(1,3)_{2}}\,,
\end{eqnarray}
where the components are labelled in terms of their SM quantum numbers, $(\su 3, \su 2)_\text{Y}$, and we indicate the intrinsic $P_1$ parity by colour: blue for even components and red for odd ones. We recall that an overall sign $\eta$ can be defined for bulk matter fields (but not for gauge ones).
It is already clear that the best candidates to contain the SM fermions are the ${\bf 5}$ and ${\bf 10}$, as they are the irreps that only contain components with SM-like quantum numbers. The main reason is that for $\su 5$, only one parity can be defined, hence it is not possible to separate states with the same parity within the same multiplets. For instance, a zero mode in the $(\bar{3},2)_{-1/6}$ in the ${\bf 40}$ will necessarily be accompanied by zero modes $(8,2)_{-1/2}$ and $(3,2)_{7/6}$. The only exception is the symmetric S, which may contain the quark doublet $q_L \equiv (3,2)_{1/6}$.

Having at our disposal only $P_1$ and the identity $P_0$, the SM as 4D remnant can be obtained via the following two choices:  
\begin{equation}
    P_1\times P_1'\, \qquad  P_0\times P_1\,. 
\end{equation}

\subsubsection{Case $P_1 \times P_1'$:} 
As the two $P_1$ parities need to be aligned to preserve the SM remnant, this case would be equivalent to an orbifold with a single parity, $S^1/\mathbb{Z}_2$. As a consequence, for bulk fermions, all components will feature a zero mode, either left-handed or right-handed. This is a source of many unwanted states. For instance, a bulk ${\bf 5}$ with parities $(\eta_1, \eta_1')=(-,-)$ will have a right-handed zero mode $(3,1)_{-1/3} \equiv d_R$ but also a left-handed $(1,2)_{1/2}$, which has conjugate quantum numbers to the lepton doublets $l_L$. Hence, this possibility is not viable.~\footnote{We refer the reader to Appendix~\ref{app:BCs} for a possible way to salvage this case, leading however to non-minimal models.}

\subsubsection{Case $P_0 \times P_1$:} 
In this case, the parity $P_0$ allows to remove the zero modes for some of the multiplet components. The minimal choice for the SM fermion embedding relies on ${\bf 5}$ and ${\bf 10}$, as follows:
\begin{equation}
    \Psi_5^{(-,-)} \supset d_R\,, \quad \Psi_{\bar{5}}^{(+,-)} \supset l_L\,, \quad \Psi_{10}^{(+,-)} \supset q_L\,, \quad \Psi_{\bar{10}}^{(-,-)} \supset u_R + e_R\,,
\end{equation}
where the superscript indicates the parities $(\eta_0, \eta_1)$ for the fields. This set corresponds to the model in Ref.~\cite{Cacciapaglia:2020qky}. Yukawa couplings can be added in the bulk via a scalar $\Phi_5^{(-,-)}\supset \varphi_H$ containing only a SM-like Brout-Englert-Higgs doublet (Higgs doublet for short) as zero mode. The Yukawas can be written as \cite{Cacciapaglia:2020qky,Cacciapaglia:2022nwt}:
\begin{equation} \label{eq:Yuksu5}
    \mathcal{L}_\text{Yuk} = - Y_u\ \overline{\Psi}_{\bar{10}} \Psi_{10} \Phi_{5} - Y_d \ \overline{\Psi}_{10} \Psi_{5} \Phi_5 - Y_l\ \overline{\Psi}_{\bar{5}} \Psi_{\bar{10}} \Phi_5 + \text{h.c.}
\end{equation}
For this model, we obtain the following 5D beta function for $n_g$ copies of the bulk fields
\begin{equation}
    b_5 = \frac{4}{3} (13-4 n_g)\,,
\end{equation}
hence allowing for a fixed point for up to three bulk generations, $n_g \leq 3$.

The quark doublet $q_L$ can also stem from the symmetric, hence replacing $\Psi_{10}$ with $\Psi_{15}^{(+,-)} \supset q_L$ in the previous set. As a consequence, it is not possible to write the up-type Yukawa with the Higgs within the ${\bf 5}$. However, a second Higgs doublet can be introduced via a ${\bf \bar{45}}$: $\Phi_{\bar{45}}^{(+,-)} \supset \varphi_H' + (8,2)_{1/2} + (\bar{3},2)_{-7/6}$. The Yukawas now read:
\begin{equation}
    \mathcal{L}_\text{Yuk} = - Y_t \ \overline{\Psi}_{\bar{10}} \Psi_{15} \Phi_{\bar{45}} - Y_d\ \overline{\Psi}_{15} \Psi_5 \Phi_5 - Y_l \overline{\Psi}_{\bar{5}} \Psi_{\bar{10}} \Phi_5 - Y_l' \ \overline{\Psi}_{\bar{5}} \Psi_{\bar{10}} \Phi_{\bar{45}}^\ast + \text{h.c.}
\end{equation}
The 4D model, therefore, consists of a two-Higgs-doublet model (2HDM) \cite{Branco:2011iw}, with additional coloured scalars. Depending on the lepton Yukawas, we can have a type-II 2HDM for $Y_l' = 0$ or a flipped (Y) 2HDM for $Y_l = 0$. For this model, the 5D beta function for $n_g$ copies of the bulk fermions reads
\begin{equation}
    b_5 = \frac{40}{3} - 8 n_g\,,
\end{equation}
hence allowing strictly one bulk generation, $n_g = 1$.

Finally note that GHU cannot be achieved, as the adjoint lacks a Higgs doublet, and SUSY is not necessary as well.

\subsection{SU(6) (rank 5)} \label{subsec:su6}

For SU(6), the only remnant 4D gauge group that contain the SM is
\begin{equation} \label{eq:remnant6}
    \mathcal{G}_\text{4D} = \mathcal{G}_\text{SM} \times \uu_Z\,.
\end{equation}
We can define three non-trivial parities such that $\mathcal{H}_i \supset \mathcal{G}_\text{SM}$:
\begin{eqnarray}
    \mathcal{P}_1 &=& \text{diag} (+,\ +,\ +,\ +,\ +,\ -) \quad \mathcal{H}_1 = SU(5)\times U(1)_{Z1}\,, \\
    \mathcal{P}_2 &=& \text{diag} (+,\ +,\ +,\ -,\ -,\ +) \quad \mathcal{H}_2 = SU(4)\times SU(2)\times U(1)_{Z2}\,, \\
    \mathcal{P}_3 &=& \text{diag} (+,\ +,\ +,\ -,\ -,\ -) \quad \mathcal{H}_3 = SU(3)\times SU(3)\times U(1)_{Z3}\,.
\end{eqnarray}
The parity matrices $\mathcal{P}_i$ are aligned in such a way that all intersections of the three parities lead to the remnant in Eq.~\eqref{eq:remnant6}, henceforth we will consider models based on $P_1 \times P_2$, $P_1\times P_3$ and $P_2 \times P_3$.

Under $P_1$, the relevant $\su 6$ irreps decompose as
\begin{eqnarray}
        \text{Adj} = {\bf 35} &=& \tblu{(24)_{0}} \oplus \tblu{(1)_{0}} \oplus \tred{(5)_{3}} \oplus \tred{(\bar{5})_{-3}}\,, \\
        \text{F} = {\bf 6} &=& \tblu{(5)_{1/2}} \oplus \tred{(1)_{-5/2}} \\
        \text{A} = {\bf 15} &=& \tblu{(10)_{1}} \oplus \tred{(5)_{-2}}\,, \\
        \text{S} = {\bf 21} &=& \tblu{(15)_{1}} \oplus \tblu{(1)_{-5}} \oplus \tred{(5)_{-2}}\,, \\
        \text{A}_3 = {\bf 20} &=& \tblu{(\bar{10})_{3/2}} \oplus \tred{(10)_{-3/2}}\,,
\end{eqnarray}
where the component quantum numbers are indicated as $(\su 5)_{Q_{Z1}}$ with
\begin{equation}
        Q_{Z1} = \frac{1}{2} \text{diag} (1,\ 1,\ 1,\ 1,\ 1,\ -5)\,,
    \end{equation}
and the colours label their intrinsic parities. As $\su 5 \supset \mathcal{G}_\text{SM}$, the decomposition above already gives meaningful insights on the model building in the fermionic sector. Interestingly, a set of bulk ${\bf 15} + {\bf \bar{15}}$ contains all the fields in the $\su 5$ model of Ref.~\cite{Cacciapaglia:2020qky}. The presence of an extra $\uu$ could allow to realise a flipped $\su 5$ embedding, where one generation of SM fermions is contained in $(10)_1 + (\bar{5})_{-3} + (1)_{5}$. From the charge assignments, this would require to add in the bulk ${\bf 35} + {\bf 15} + {\bf \bar{15}} + {\bf \bar{21}}$: the main price to pay would be the presence of many unwanted chiral zero modes, hence we will not consider this possibility further. Henceforth, we can define the remnant $\uu$ charges as:
\begin{equation} \label{eq:remnantU1}
     Y = \text{diag} \left( -\frac{1}{3},\ -\frac{1}{3},\ -\frac{1}{3},\ \frac{1}{2},\ \frac{1}{2},\ 0 \right)\,, \qquad Q_Z \equiv Q_{Z1}\,.
\end{equation}

Under $P_2$, the decomposition in $(\su 4, \su 2)_{Q_{Z2}}$ reads:
    \begin{eqnarray}
        {\bf 35} &=& \tblu{(15,1)_{0}} \oplus \tblu{(1,3)_{0}} \oplus \tblu{(1)_{0}} \oplus \tred{(4,2)_{3}} \oplus \tred{(\bar{4},2)_{-3}}\,, \\
        {\bf 6} &=& \tblu{(4,1)_{1}} \oplus \tred{(1,2)_{-2}} \\
        {\bf 15} &=& \tblu{(6,1)_{2}} \oplus \tblu{(1,1)_{-4}} \oplus \tred{(4,2)_{-1}}\,, \\
        {\bf 21} &=& \tblu{(10,1)_{2}} \oplus \tblu{(1,3)_{-4}} \oplus \tred{(4,2)_{-1}}\,, \\
        {\bf 20} &=& \tblu{(\bar{4},1)_{3}} \oplus \tred{(6,2)_{0}} \oplus \tblu{(4,1)_{-3}}\,,
    \end{eqnarray}
where
\begin{equation}
    Q_{Z2} = \text{diag} (1,\ 1,\ 1,\ -2,\ -2,\ 1)\,,
\end{equation}
and the colours indicate the intrinsic parity. In this case, $\su 2$ is the weak group, while $\su 4 \supset \su 3 \times \uu_4$, with
\begin{equation}
        Q_4 = \text{diag} (1,\ 1,\ 1,\ 0,\ 0,\ -3)\,.
\end{equation}
Matching to the definitions of the remnand $\uu$'s in Eq.~\eqref{eq:remnantU1}, we see that
\begin{equation}
    Q_{Z} = \frac{3 Q_4 - Q_{Z2}}{4}\,, \quad Y = - \frac{Q_4 + 3 Q_{Z2}}{12} \,.
\end{equation}

Under $P_3$, the decomposition in $(\su 3, \su 3)_{Q_{Z3}}$ components reads:
    \begin{eqnarray}
        {\bf 35} &=& \tblu{(8,1)_{0}} \oplus \tblu{(1,8)_{0}} \oplus \tblu{(1)_{0}} \oplus \tred{(3,\bar{3})_{2}} \oplus \tred{(\bar{3},3)_{-2}}\,, \\
        {\bf 6} &=& \tblu{(3,1)_{1}} \oplus \tred{(1,3)_{-1}} \\
        {\bf 15} &=& \tblu{(\bar{3},1)_{2}} \oplus \tblu{(1,\bar{3})_{-2}} \oplus \tred{(3,3)_{0}}\,, \\
        {\bf 21} &=& \tblu{(6,1)_{2}} \oplus \tblu{(1,6)_{-2}} \oplus \tred{(3,3)_{0}}\,, \\
        {\bf 20} &=& \tblu{(1,1)_{3}} \oplus \tred{(\bar{3},3)_{1}} \oplus \tblu{(3,\bar{3})_{-1}} \oplus \tred{(1,1)_{-3}}\,.
    \end{eqnarray}
where 
\begin{equation}
    Q_{Z3} = \text{diag} (1,\ 1,\ 1,\ -1,\ -1,\ -1)\,,
\end{equation}
and the colours label the parities. Now, it is the first $\su 3$ that reproduces the SM one, while the second $\su 3 \supset \su 2 \times \uu_3$, with 
    \begin{equation}
        Q_3 = \text{diag} (0,\ 0,\ 0,\ 1,\ 1,\ -2)\,.
    \end{equation}
Matching with Eq.~\eqref{eq:remnantU1} gives:
\begin{equation}
 Q_{Z1} = \frac{2 Q_3 + Q_{Z3}}{2}\,, \qquad Y = \frac{Q_3-2 Q_{Z3}}{6} \,.   
\end{equation}    

Before proceeding with a systematic analysis of the fermion embedding, it is useful to compare the symmetry breaking patterns to the $\su 5$ case studied above. For the cases $P_1 \times P_2$ and $P_1 \times P_3$, the $\su 5$ subgroup is broken in a similar way as in the models with $\su 5/P_0\times P_1$: what distinguishes the two $\su 6$ models is how various sub-components of the $\su 6$ irreps receive their intrinsic parity. In the case $P_2 \times P_3$, the $\su 5$ subgroup is broken in the same way by the two parities, hence being equivalent to the $\su 5/P_1 \times P_1'$ model. Following the same arguments we used in $\su 5$, this case can be ruled out.

For our choice of alignment, all combinations of parities lead to the same embedding of the SM within $\su 6$, hence the intrinsic parities under $P_{1,2,3}$ can be listed on the same components as shown in Table~\ref{tab:su6par} for the relevant irreps.

\begin{table}[h!]
\centering
\begin{tabular}{|c|c|c|c|}
\hline
Adjoint (${\bf 35}$) & $P_1$ & $P_2$ & $P_3$ \\ \hline \hline
$(8,1)_{0,0}$ & \even & \even & \even\\
$(1,3)_{0,0}$ & \even & \even & \even\\
$(1,1)_{0,0}$ & \even & \even & \even\\
$(3,2)_{-5/3,0}$ & \even & \odd  & \odd \\
$(\bar{3},2)_{5/3,0}$ & \even & \odd & \odd \\\hline
$(1,1)_{0,0}$ & \even & \even & \even \\ \hline
$(3,1)_{-1/3,3}$ & \odd & \even & \odd \\
$(1,2)_{1/2,3}$ & \odd & \odd & \even \\ \hline
$(\bar{3},1)_{1/3,-3}$ & \odd & \even & \odd\\
$(1,2)_{-1/2,-3}$ & \odd & \odd & \even \\
\hline
\end{tabular} 
\hfill
\begin{tabular}{c}
\begin{tabular}{|c|c|c|c|}
\hline
F (${\bf 6}$) & $P_1$ & $P_2$ & $P_3$ \\ \hline \hline
$(3,1)_{-1/3,1/2}$ & \even & \even & \even\\
$(1,2)_{1/2,1/2}$ & \even & \odd & \odd\\\hline
$(1,1)_{0,-5/2}$ & \odd & \even & \odd \\\hline
\end{tabular} \\
\\
\\
\begin{tabular}{|c|c|c|c|}
\hline
S (${\bf 21}$) & $P_1$ & $P_2$ & $P_3$ \\ \hline \hline
$(3,2)_{1/6,1}$ & \even & \odd & \odd\\
$(6,1)_{-2/3,1}$ & \even & \even & \even\\
$(1,3)_{1,1}$ & \even & \even & \even \\ \hline
$(1,1)_{0,-5}$ & \even & \even & \even \\ \hline
$(3,1)_{-1/3,-2}$ & \odd & \even & \odd\\
$(1,2)_{1/2,-2}$ & \odd & \odd & \even\\\hline
\end{tabular}
\end{tabular}
\hfill
\begin{tabular}{c}
\begin{tabular}{|c|c|c|c|}
\hline
A (${\bf 15}$) & $P_1$ & $P_2$ & $P_3$ \\ \hline \hline
$(3,2)_{1/6,1}$ & \even & \odd & \odd\\
$(\bar{3},1)_{-2/3,1}$ & \even & \even & \even\\
$(1,1)_{1,1}$ & \even & \even & \even \\ \hline
$(3,1)_{-1/3,-2}$ & \odd & \even & \odd\\
$(1,2)_{1/2,-2}$ & \odd & \odd & \even\\\hline
\end{tabular}  \\
\\
\begin{tabular}{|c|c|c|c|}
\hline
A$_3$ (${\bf 20}$) & $P_1$ & $P_2$ & $P_3$ \\ \hline \hline
$(\bar{3},2)_{-1/6,3/2}$ & \even & \odd & \odd\\
$(3,1)_{2/3,3/2}$ & \even & \even & \even\\
$(1,1)_{-1,3/2}$ & \even & \even & \even \\ \hline
$(3,2)_{1/6,-3/2}$ & \odd & \odd & \even\\
$(\bar{3},1)_{-2/3,-3/2}$ & \odd & \even & \odd\\
$(1,1)_{1,-3/2}$ & \odd & \even & \odd \\\hline
\end{tabular} 
\end{tabular}
\caption{\label{tab:su6par} Intrinsic parities of the SM components of the adjoint (${\bf 35}$), relevant for the gauge field, the fundamental F (${\bf 6}$), the anti-symmetric A (${\bf 15}$), the symmetric S (${\bf 21}$), and the three-index anti-symmetric A$_3$ (${\bf 20}$). The U(1) charges refer to $Y$ and $Q_{Z}$. The horizontal lines separate multiplets of the subgroup $\su 5$ that contains the SM gauge symmetry.}
\end{table}

\subsubsection{Case $P_1\times P_2$:}

This is the most interesting symmetry breaking pattern as the gauge sector contains a Higgs doublet at the zero mode level, as shown in Table~\ref{tab:su6par} for the adjoint. Hence, at least some of the Yukawa couplings can emerge via the GHU mechanism and inherit the UV fixed point from the gauge running.

To construct a viable fermion embedding, we first list all possible bulk fermions with parity assignments leading to only SM-like zero modes:
\begin{eqnarray}
    && {\bf 6}^{(-,-)} \supset d_R\,, \;\; {\bf \bar{6}}^{(+,-)} \supset l_L + \nu_R\,, \;\; {\bf 15}^{(+,-)} \supset q_L + d_R\,, \;\; {\bf \bar 15}^{(-,-)} \supset l_L + u_R + e_R\,, \nonumber\\
    && {\bf 21}^{(+,-)} \supset q_L + d_R\,, \;\; {\bf 20}^{(-,-)} \supset q_L + u_R + e_R\,.
\end{eqnarray}
Viable models can be found by combining the irreps above to form a complete SM generation.

Hence, a first viable model is characterised by
\begin{equation}
    \Psi_{15}^{(+,-)} \supset q_L + d_R\,, \quad \Psi_{\bar{15}}^{(-,-)} \supset l_L + u_R + e_R\,, \quad  \Phi_{15}^{(-,-)} \supset \varphi_H'\,.
\end{equation}
While down and lepton Yukawas are generated by GHU, the top Yukawa is due to the bulk scalar:
\begin{equation}
    \mathcal{L}_\text{Yuk} = - Y_u\ \overline{\Psi}_{\bar{15}} \Psi_{15} \Phi_{15} + \text{h.c.}
\end{equation}
so that the remnant 4D theory consists of a type-II 2HDM. The beta function reads
\begin{equation}
    b_5 = \frac{61-16 n_g}{3}\,,
\end{equation}
hence supporting up to 3 generations in the bulk.

Another possibility is to obtain the quark doublet from the A$_3$ irrep, leading to:
\begin{equation}
    \Psi_{20}^{(-,-)} \supset q_L + u_R + e_R\,, \quad \Psi_{6}^{+,+} \supset d_R\,, \quad \Psi_{\bar 6}^{(+,-)} \supset l_L + \nu_R\, \quad \Phi_{15}^{(-,-)}\supset \varphi_H'\,.
\end{equation}
While up Yukawa stems from GHU (together with a neutrino Yukawa), the other two can be written as:
\begin{equation}
    \mathcal{L}_\text{Yuk} = - Y_d\ \overline{\Psi}_{20} \Psi_{6} \Phi_{15} - Y_l\ \overline{\Psi}_{\bar 6} \Psi_{20} \Phi_{15} + \text{h.c.}
\end{equation}
so that the remnant 4D theory consists of a type-II 2HDM with right-handed neutrinos.
The beta function read
\begin{equation}
    b_5 = \frac{61-16 n_g}{3}\,,
\end{equation}
hence also supporting up to three bulk generations.

In principle, a variant of the first model can be constructed by replacing the ${\bf 15}$ with a symmetric ${\bf 21}$, $\Psi_{21}^{(+,-)} \supset q_L + d_R$. However, the up quark Yukawa would need a Higgs embedded in a ${\bf 105}$ of $\su 6$, hence generating many non-SM zero mode scalars. The beta function of this model reads
\begin{equation}
    b_5 = \frac{37}{3} - 8 n_g\,,
\end{equation}
hence allowing only one bulk generation.

\subsubsection{Case $P_1\times P_3$:}

For this symmetry breaking pattern, the gauge scalars contain a zero mode with quantum numbers $(3,1)_{-1/3}$ that should not be allowed to develop a vacuum expectation value. This poses a first challenge for model building in this case.

The bulk irreps containing SM-like zero modes are now:
\begin{eqnarray}
    {\bf 6}^{(-,-)} \supset d_R + \nu_R^c\,, \;\; {\bf \bar{6}}^{(+,-)} \supset l_L\,, \;\; {\bf 15}^{(+,-)} \supset q_L + l_L^c\,, \;\; {\bf \bar{15}}^{(-,-)} \supset u_R + e_R + d_R^c\,, \;\; {\bf 21}^{(+,-)} \supset q_L + l_L^c\,.
\end{eqnarray}
Hence, the only viable combination stems from the anti-symmetric irreps. However, as the chirality of the down quark does not match that of the quark doublets (they are both left-handed) as well as for charge leptons (they are both right-handed), it will not be possible to write bulk Yukawas for them. In fact, due to the vector-like nature of the 5D spinors, bulk masses and Yukawas can only couple field with opposite 4D chirality.
The only viable model, therefore, reads
\begin{equation}
    \Psi_{15}^{(+,-)} \supset q_L + l_L^c\,, \quad \Psi_{\bar{15}}^{(-,-)} \supset u_R + e_R + d_R^c\,, \quad  \Phi_{15}^{(-,+)} \supset \varphi_H'\,.
\end{equation}
The only bulk Yukawa gives mass to the up quarks:
\begin{equation}
    \mathcal{L}_\text{Yuk} = - Y_u\ \overline{\Psi}_{\bar{15}} \Psi_{15} \Phi_{15} + \text{h.c.}
\end{equation}
while the other fermions remain massless. The origin for their Yukawa couplings may be in localised interactions, and we will investigate their feasibility in the next section.
As before, the ${\bf 15}$ can be replaced by the symmetric ${\bf 21}$ at the price of introducing the Higgs via a ${\bf 105}$. The beta functions are the same as in the $P_1\times P_2$ cases.

\subsubsection{Case $P_2\times P_3$:}

In this case, the gauge scalars contain a zero mode $(3,2)_{-5/3}$. However, a quick inspection of the intrinsic parity assignments in Table~\ref{tab:su6par} reveals that for all irreps, when selecting a component with SM-like zero mode, it is accompanied by a zero mode with conjugate quantum numbers. This is due to the fact that states within the same $\su 5$ multiplet have opposite parities, as for the case of $\su 5/P_1\times P_1'$. Hence, this case is ruled out.

\subsection{Larger groups}

For a generic $\su{N}$ bulk group, the parities $P_i$ can break it as $\su{N}  \to \su{m} \times \su{N-m} \times \uu$. The intersection of the two diagonal parities preserves at most 4 SU gauge factors and a suitable number of $\uu$'s. Hence, the SM gauge symmetry as a remnant can be obtained in two distinct ways:
\begin{eqnarray}
    \mathcal{G}_\text{4D} &=& \su 3 \times \su 2 \times \uu_Y \times \su{K} \times \uu_X\,, \\
    \mathcal{G}_\text{4D} &=& \su 3 \times \su 2 \times \uu_Y \times \su{K-m} \times \su{m} \times \uu_{X1} \times \uu_{X2}\,,
\end{eqnarray}
where $K = N-5$ and $m$ is any integer $2 \leq m \leq K-2$. From the $\su 6$ analysis we also learned that the SM zero modes can always be considered as stemming from an $\su 5$ subgroup, and this fact remains true for larger groups.

\subsubsection{$\mathcal{G}_\text{4D} = \mathcal{G}_\text{SM} \times \su{N-5} \times \uu_X$ :}

This case can be achieved by the intersection of two parities out of the three listed below:
\begin{eqnarray}
    \mathcal{P}_1 &=& \text{diag} (+,\ +,\ +,\ +,\ +,\ -, \dots -) \quad \mathcal{H}_1 = \su{5}\times \su{N-5} \times \uu_{Z1}\,,  \\
    \mathcal{P}_2 &=& \text{diag} (+,\ +,\ + ,\ -,\ -,\ +,\ \dots +) \quad \mathcal{H}_2 = \su{2}\times \su{N-2} \times \uu_{Z2}\,, \\
     \mathcal{P}_3 &=& \text{diag} (+,\ +,\ +,\ -,\ -,\ -,\ \dots -) \quad \mathcal{H}_3 = \su{3}\times \su{N-3} \times \uu_{Z3}\,.
\end{eqnarray}
This is a straightforward generalisation of the $\su 6$ cases, where the SM zero modes are organised in terms of the $\su 5$ subgroup spanning the first 5 components of the gauge indices. As such, it is instructive to first look at the decompositions of the relevant irreps in terms of $\mathcal{H}_1$:
\begin{eqnarray}
        \text{Adj} &=& \tblu{(24,1)_{0}} \oplus \tblu{(1,1)_{0}} \oplus \tblu{(1,\text{Adj})_{0}} \oplus \tred{(5, \bar{\text{F}})_{N}} \oplus \tred{(\bar{5}, \text{F})_{-N}}\,, \\
        \text{F} &=& \tblu{(5,1)_{N-5}} \oplus \tred{(1,\text{F})_{-5}} \\
        \text{A} &=& \tblu{(10,1)_{2N-10}} \oplus \tred{(5,\text{F})_{N-10}} \oplus \tblu{(1,\text{A})_{-10}}\,, \\
        \text{S} &=& \tblu{(15,1)_{2N-10}} \oplus \tred{(5,\text{F})_{N-10}} \oplus \tblu{(1,\text{S})_{-10}}\,, \\
        \text{A}_3 &=& \tblu{(\bar{10},1)_{3 N-15}} \oplus \tred{(10,\text{F})_{2N-15}} \oplus \tblu{(5,\text{A})_{N-15}} \oplus \tred{(1,\text{A}_3)_{-15}}\,, \\
        \text{A}_4 &=& \tblu{(\bar{5},1)_{4 N-20}} \oplus \tred{(\bar{10},\text{F})_{3N-20}} \oplus \tblu{(10,\text{A})_{2N-20}} \oplus \tred{(5,\text{A}_3)_{N-20}} \oplus \tblu{(1,\text{A}_4)_{-20}}\,,
\end{eqnarray}
where
\begin{equation}
    Q_{Z1} = \text{diag} (K,\ K,\ K,\ K,\ K,\ -5,\ \dots -5)\,, \;\; \text{with} \;\; K=N-5\,.
\end{equation}
The breaking of $\su 5$ implies that components inside the same irrep of $\su 5$ will have different overall parity assignments. Hence, it is inevitable to have zero modes belonging to states that are also multiplets of $\su{N-5}$: this gives them a multiplicity that is contrary to our assumption of minimality. In other words, it is impossible to obtain the SM zero modes from singlets of $\su{N-5}$ without also having zero modes belonging to multiplets of $\su{N-5}$. Hence, no minimal aGUT model can be constructed along this path.

\subsubsection{$\mathcal{G}_\text{4D} = \mathcal{G}_\text{SM} \times \su{N-5-m} \times \su{m} \times \uu_{X_1} \times \uu_{X2}$ :}

This case is closely related to the previous one, as it can be obtained by the same parities $P_2$ and $P_3$ where some of the ``$-$'' in the last $N-5$ indices are turned into ``$+$''. Hence, in the above decomposition, the multiplets of $\su{N-5}$ are split. Nevertheless, it remain the problem than unwanted zero modes always appear, which carry SM quantum numbers. Hence, no minimal aGUT can be constructed.

\vspace{0.5cm}

The combined analysis of these two symmetry breaking patterns shows that no minimal aGUT models can stem from models based on SU(N) with $N \geq 7$.

\subsection{Summary}

In summary, our exploration of the SU(N) kinship yielded 7 viable models at this stage, 2 stemming from $\su 5$ and 5 from $\su 6$. The main ingredients are listed in Table~\ref{tab:models}. Only 3 models, however, feature a minimal Higgs sector, containing either one or two Higgs doublets. The other models always feature coloured scalar zero modes, which must not develop any vacuum expectation value, hence imposing further challenges at low energies.

\begin{table}[htb]
\centering
\begin{tabular}{|c|c|c|c|c|c|c|c|}
\hline
Name & $\mathcal{G}_\text{bulk}$ & Fermions & Scalars & Yukawas & $n_g$ bulk & Higgs sector & Minimal? \\ \hline \hline
\multicolumn{8}{|c|}{$\mathcal{G}_\text{4D} = \mathcal{G}_\text{SM}$} \\ 
\hline
\multirow{2}{*}{5A} & \multirow{2}{*}{$\su 5$} & $\Psi_5 \supset d_R\,, \;\; \Psi_{\bar{5}} \supset l_L\,,$ & \multirow{2}{*}{$\Phi_5 \supset \varphi_H$} & \multirow{2}{*}{All bulk} & \multirow{2}{*}{$\leq 3$} & \multirow{2}{*}{SM-like} & \multirow{2}{*}{Yes} \\
 & & $\Psi_{10} \supset q_L\,, \;\; \Psi_{\bar{10}} \supset u_R + e_R$ & & & & & \\
\hline
\multirow{2}{*}{5S} & \multirow{2}{*}{$\su 5$} & $\Psi_5 \supset d_R\,, \;\; \Psi_{\bar{5}} \supset l_L\,,$ & $\Phi_5 \supset \varphi_H$ & \multirow{2}{*}{All bulk} & \multirow{2}{*}{$1$} & 2HDM & \multirow{2}{*}{No} \\
 & & $\Psi_{15} \supset q_L\,, \;\; \Psi_{\bar{10}} \supset u_R + e_R$ & $\Phi_{\bar{45}} \supset \varphi_H' + \dots$ & & & Type-II or flip & \\
\hline
\multicolumn{8}{|c|}{$\mathcal{G}_\text{4D} = \mathcal{G}_\text{SM} \times \uu_Z$} \\ 
\hline
\multirow{2}{*}{6A} & \multirow{2}{*}{$\su 6$} & $\Psi_{15} \supset q_L + d_R\,,$ & $\text{Adj} \supset \varphi_H$ & $d, e$ GHU & \multirow{2}{*}{$\leq 3$} & 2HDM & \multirow{2}{*}{Yes} \\
 & & $\Psi_{\bar{15}} \supset l_L + u_R + e_R$ & $\Phi_{15} \supset \varphi_H'$ & $u$ bulk & & Type-II & \\
\hline
\multirow{2}{*}{6A flip} & \multirow{2}{*}{$\su 6$} & $\Psi_{20} \supset q_L + u_R + e_R\,,$ & $\text{Adj} \supset \varphi_H$ & $u$ GHU & \multirow{2}{*}{$\leq 3$} & 2HDM & \multirow{2}{*}{Yes} \\
 & & $\Psi_{6} \supset d_R\,, \;\; \Psi_{\bar{6}} \supset l_L + \nu_R$ & $\Phi_{15} \supset \varphi_H'$ & $d, e$ bulk & & Type-II & \\
\hline
\multirow{2}{*}{6S} & \multirow{2}{*}{$\su 6$} & $\Psi_{21} \supset q_L + d_R\,,$ & $\text{Adj} \supset \varphi_H$ & $d, e$ GHU & \multirow{2}{*}{$1$} & 2HDM & \multirow{2}{*}{No} \\
 & & $\Psi_{\bar{15}} \supset l_L + u_R + e_R$ & $\Phi_{105} \supset \varphi_H' + \dots$ & $u$ bulk & & Type-II & \\
\hline
\multirow{2}{*}{6A'} & \multirow{2}{*}{$\su 6$} & $\Psi_{15} \supset q_L + l_L^c\,,$ & \multirow{2}{*}{$\Phi_{15} \supset \varphi_H$} & \multirow{2}{*}{$u$ bulk} & \multirow{2}{*}{$\leq 3$} & \multirow{2}{*}{SM-like} & \multirow{2}{*}{No} \\
 & & $\Psi_{\bar{15}} \supset u_R + e_R + d_R^c$ &  &  & &  & \\
\hline
\multirow{2}{*}{6S'} & \multirow{2}{*}{$\su 6$} & $\Psi_{21} \supset q_L + l_L^c\,,$ & \multirow{2}{*}{$\Phi_{105} \supset \varphi_H + \dots$} & \multirow{2}{*}{$u$ bulk} & \multirow{2}{*}{$1$} & \multirow{2}{*}{SM-like} & \multirow{2}{*}{No} \\
 & & $\Psi_{\bar{15}} \supset u_R + e_R + d_R^c$ &  &  & &  & \\
\hline
\end{tabular}
\caption{\label{tab:models} Summary of the viable models in the SU(N) kinship. The second and third column indicate the embedding of the SM fermions and scalars, where Adj indicates the gauge multiplet (gauge-scalars). The origin of the Yukawa couplings is indicated in the fifth column, while the sixth indicate the maximum number of bulk SM generations admitted by the gauge UV fixed point. Finally, the last two columns qualify the SM Higgs sector. We consider as minimal sectors containing a single or two Higgs doublets, while ``No'' indicates the presence of additional (coloured) scalar zero modes. }
\end{table}

It is also noteworthy that only 6A flip contains right-handed neutrinos, with Yukawas generated via GHU. In all other models, except 6S and 6S', right-handed neutrinos can be added as singlets of the bulk gauge symmetry and Yukawa with the bulk scalar. Finally, the models with the symmetric irrep, 5S, 6S and 6S', do not allow for 3 bulk generations.  

The next step in the feasibility analysis consists in checking the UV fate of the Yukawa couplings. We recall that couplings generated by GHU inherit the gauge fixed point, hence they naturally unify with the bulk gauge couplings. Nevertheless, all models have bulk Yukawa couplings that could run into Landau poles in 5D. Furthermore, additional generations and/or Yukawa couplings may be generated on the boundary, if compatible with the boundary gauge symmetry. We will process this last and final validation of the models in the next section.

\section{The final test: Yukawa sector in the UV} \label{sec:Yukawa}

The Yukawa sector poses serious challenges to the aGUT paradigm, as their running may not feature a fixed point but run into a Landau pole right above the compactification scale. This would induce a low cut-off for the 5D theory. Furthermore, even if the fixed point existed, it may not be attractive, hence limiting the values of the Yukawa couplings at low energies to values that are not compatible with the SM. An analysis for the 5A model has been presented in Ref.~\cite{Cacciapaglia:2020qky}, and we will simply recap the main results below, together with the analogous analysis for the other models. Note that here we focus on finding models with fixed points for all Yukawa couplings, leaving a numerical analysis of the low energy behaviour and predictions to further investigation.

In general, for a single Yukawa coupling, one can write an analogue to the running equation ~\eqref{eq:5Derg} in terms of the 't Hooft coupling $\tilde{\alpha}_y$, defined as in Eq.~\eqref{eq:alphatildedef}, as
\begin{equation}
    2 \pi \frac{d}{d \ln \mu} \tilde{\alpha}_y = 2 \pi \tilde{\alpha}_y + c_{y} \tilde{\alpha}_y^2 - d_{y} \tilde{\alpha} \tilde{\alpha}_y\,, 
\end{equation}
where $\tilde{\alpha}$ is the bulk gauge coupling and, typically, $c_{y}$ and $d_y$ are positive numbers. This equation shows two points:
\begin{itemize}
    \item[i)] The UV fixed point is repulsive as long as $c_y > 0$.
    \item[ii)] A fixed point exists iff $d_y \tilde{\alpha}^\ast > 2 \pi$, as
    \begin{equation}
        \tilde{\alpha}_y^\ast = \frac{d_y \tilde{\alpha}^\ast - 2 \pi}{c_y}\,,
    \end{equation}
    is the zero of the renormalisation equation in the UV, where the gauge coupling has reached its own fixed point value $\tilde{\alpha}^\ast$.
\end{itemize}
Because of the former, the flow to the fixed point is not guaranteed for all values of the low energy Yukawa, and the model has some predictive power. The limitation stands in the fact that non-gauge couplings among KK modes enter in the running near the KK mass, and their values are not properly calculable. Hence, any low energy prediction will be marred by this uncertainty \cite{Cacciapaglia:2020qky}. 

In models where several Yukawas appear in the bulk, a set of coupled running equations need to be solved, in the general form:
\begin{equation}
    2 \pi \frac{d}{d \ln \mu} \tilde{\alpha}_y = \left( 2 \pi  + \sum_{y'}  c_{yy'} \tilde{\alpha}_{y'} - d_y \tilde{\alpha} \right) \tilde{\alpha}_y\,,
\end{equation}
where $y,y'$ run over the allowed Yukawas (for instance, $y,y' = {u, d, e, \nu}$). In this case, the zeros are given by
\begin{equation} \label{eq:sol1}
    \tilde{\alpha}_y^\ast = \sum_{y'} c_{yy'}^{-1}\ (d_{y'} \tilde{\alpha}^\ast - 2 \pi)\,, 
\end{equation}
where $c_{yy'}^{-1}$ is the inverse of the matrix $c_{yy'}$. Details about the computation of the coefficients are given in Appendix~\ref{app:Yukawas}, with master formulae extracted and adapted from Ref.\cite{Cot2021}.

\subsection{Model 5A on $P_0\times P_1$}

This model has been described in details in Ref. \cite{Cacciapaglia:2020qky}, hence we will here recap the main results. Firstly, the minimal field content in Table~\ref{tab:models} does not allow for neutrino masses. The model can be easily extended by two gauge singlets
\begin{equation}
    \Psi_1^{(-,-)} \supset \nu_R\,, \quad \Psi_{1'}^{(-,+)}\,,
\end{equation}
where the second has no zero modes and was introduced to generate an asymmetric Dark Matter candidate \cite{Cacciapaglia:2020qky}.
Hence, the complete set of bulk Yukawas reads:
\begin{equation} \label{eq:Yuk5A}
\mathcal{L}_\text{Yuk}^\text{5A} = - Y_u\ \overline{\Psi}_{\bar{10}} \Psi_{10} \Phi_{5} - Y_d \ \overline{\Psi}_{10} \Psi_{5} \Phi_5 - Y_l\ \overline{\Psi}_{\bar{5}} \Psi_{\bar{10}} \Phi_5 - Y_\nu\ \overline{\Psi}_1 \Psi_{\bar{5}} \Phi_5 - Y_\chi\ \bar{\Psi}_5 \Psi_{1'} \Phi_5 + \text{h.c.}
\end{equation}
While the first four generate the related Yukawas in the SM (including a Dirac Yukawa for the neutrinos), the latter is a Yukawa coupling for the Dark Matter candidate. It has been shown in Ref.~\cite{Cacciapaglia:2020qky} that the bulk Yukawa above conserve baryon number, hence no proton decay is induced. The KK states from fields with parities $(+,-)$ and $(-,+)$ feature baryon number $1/6$ or $1/2$, hence they cannot decay into SM final states and were dubbed ``Indalo-particles'': the additional singlet is required for obtaining a realistic Dark Matter candidate (being mainly gauge singlet).

Labelling the index $y = {u,\ d,\ l,\ \nu,\ \chi}$, the coefficients of the running equations are given by the following matrices:
\begin{equation}
    c_{yy'} = \begin{pmatrix}
    60 & \frac{13}{2} & \frac{13}{2} & 2 & 2 \\
    78 & \frac{11}{2} & 4 & 2 & \frac{9}{2} \\
    78 & 4 & \frac{11}{2} & \frac{9}{2} & 2 \\
    48 & 4 & 9 & 5 & 2 \\
    48 & 9 & 4 & 2 & 5 
    \end{pmatrix}\,, \quad d_y = \begin{pmatrix}
    \frac{114}{5} \\ \frac{93}{5} \\ \frac{93}{5} \\ 6 \\ 6
    \end{pmatrix}\,.
\end{equation}
Note that the normalisations for the couplings differ from the ones adopted in Ref.~\cite{Cacciapaglia:2020qky}, but this does not affect the fixed point analysis.
From Eq.~\eqref{eq:sol1}, we see that the zeros for the first three Yukawas are always positive, but the ones for the neutrino and DM ones are always negative, henceforth signalling the absence of a UV fixed point for the system. Hence the model with the most general Yukawas in the bulk is not viable.

However, as the Yukawas for neutrino and Dark Matter can be phenomenologically very small at low energies, one could explore the possibility that only the first three Yukawas run to a fixed point. This case can be obtained by neglecting the contribution of $\nu$ and $\chi$ to the running, and find the zeros for the other three couplings, leading to
\begin{equation}
    \tilde{\alpha}_u^\ast = \frac{7}{1110} \frac{10 n_g - 19}{13-4 n_g} \pi\,, \;\; \tilde{\alpha}_d^\ast = \tilde{\alpha}_l^\ast = \frac{3}{185} \frac{20 n_g + 73}{13-4 n_g} \pi\,, 
\end{equation}
Hence, a set of UV fixed points is allowed for $n_g = 2$ and $3$. For instance, for $n_g=3$, the complete fixed point of the model reads:
\begin{equation}
    \tilde{\alpha}^\ast = \frac{3\ \pi}{2}\,, \;\; \tilde{\alpha}_u^{\ast} = \frac{77\ \pi}{1110}\,, \;\; \tilde{\alpha}_d = \tilde{\alpha}_l = \frac{399\ \pi}{185}\,.
\end{equation}
As we can see, the fixed point value for the up-type Yukawa is always very small, hence it seems hard for this model to numerically support the large values of the top Yukawa at low energies. Henceforth, this case is not realistic, while being viable from the point of view of the existence of fixed points.

Finally, motivated by the largeness of the top Yukawa, one could consider that only $\tilde{\alpha}_u$ runs to a fixed point. Neglecting the contribution of the other Yukawas, assumed to be always small and subdominant, we find
\begin{equation}
    \tilde{\alpha}_u^\ast = \frac{40 n_g + 41}{300 (13-4 n_g)} \pi\,,
\end{equation}
which is always positive. However, once the gauge and top Yukawas reach the fixed points, the other couplings will start running as a positive power of the energy, hence growing too fast.

The above analysis, partly contained in Ref.~\cite{Cacciapaglia:2020qky}, shows that only a configuration with three bulk Yukawas can feature UV fixed points for all bulk couplings. However, numerical values seem to be incompatible with the large top Yukawa coupling. Hence, the 5A model is not viable due to the Yukawa sector.

Phenomenologically, there are two ways to salvage the model:
\begin{itemize}
    \item[-] The Yukawa couplings are small enough to always run to zero in the UV. This possibility needs to be studied numerically and, as shown in Ref.~\cite{Cacciapaglia:2020qky}, it implies a lower limit on $m_\text{KK}$.
    \item[-] Some of the Yukawas are localised on the boundaries of the compact space.
\end{itemize}
In this model, Yukawas and full generations can be localised on either boundary. On the $P_0$ boundary, as $\su 5$ is preserved, the SM fermions must appear in complete $\su 5$ multiplets. If they stem from bulk fields, the multiplet is completed by states without a zero mode, hence the localised Yukawas have the same structure of the bulk ones and do preserve baryon number. Instead, if a generation is localised here, the localised couplings would violate the bulk baryon number, hence the strong bounds on $m_\text{KK}$ would stem from proton decay searches. On the $P_1$ boundary, the localised couplings only respect the SM gauge symmetries, hence baryon number can be preserved. Some possibilities for localised Yukawas have been analysed in Ref.~\cite{Cacciapaglia:2020qky}.

\subsection{Model 6A on $P_1\times P_2$}

In this model, the down quark and charged lepton Yukawas are generated by the GHU mechanism, hence they inherit the attractive fixed point of the gauge coupling. Such Yukawas need not be added to the bulk.
Neutrinos and a Dark Matter candidate can be included via singlets, like in the model 5A:
\begin{equation}
    \Psi_1^{(-,-)} \supset \nu_R\,, \quad \Psi_{1'}^{(-,+)}\,.
\end{equation}
The bulk Yukawas now read:
\begin{equation} \label{eq:Yuk6A}
    \mathcal{L}_\text{Yuk} = - Y_u\ \overline{\Psi}_{\bar{15}} \Psi_{15} \Phi_{15} - Y_\nu\ \overline{\Psi}_1 \Psi_{\bar{15}} \Phi_{15} - Y_\chi\ \bar{\Psi}_{15} \Psi_{1'} \Phi_{15} + \text{h.c.}
\end{equation}
As the field content matches the one of the model 5A, once we decompose the bulk field in $\su 5$ components, this set of bulk Yukawas also conserved baryon number. Hence, the singlet $\Psi_{1'}$ is essential to generate an Indalo Dark Matter candidate, as in Ref.~\cite{Cacciapaglia:2020qky}.

Labelling $y = {u,\ \nu,\ \chi}$, the coefficients of the running equations are given by:
\begin{equation}
    c_{yy'} = \begin{pmatrix}
    144 & \frac{1}{2} & \frac{1}{2}  \\
    24 & 10 & 2  \\
    24 & 2 & 10 
    \end{pmatrix}\,, \quad d_y = \begin{pmatrix}
    28 \\ \frac{35}{3} \\ \frac{35}{3}
    \end{pmatrix}\,.
\end{equation}
Note that the contribution of down and lepton Yukawas is included in the gauge contribution by construction.
The zeros now correspond to
\begin{equation}
    \tilde{\alpha}_u^\ast = \frac{151+88 n_g}{426 (61-16 n_g)} \pi\,,\;\; \tilde{\alpha}_\nu^\ast = \tilde{\alpha}_\chi^\ast = \frac{2 (80 n_g - 179)}{71 (61-16 n_g)} \pi\,.
\end{equation}
Hence, positive solutions only exist for $n_g = 3$, for which we find 
\begin{equation}
    \tilde{\alpha}^\ast = \frac{6\ \pi}{13}\,, \;\; \tilde{\alpha}_u^\ast = \frac{415\ \pi}{5538}\,, \;\; \tilde{\alpha}_\nu^\ast = \tilde{\alpha}_\chi^\ast = \frac{122\ \pi}{923}\,.
\end{equation}
This implies that an aGUT model, including Yukawas, can only exist for 3 fermion generations in the bulk.

It should be also noted that, removing the bulk singlets, the up Yukawa alone also has a fixed point, given by:
\begin{equation}
    \tilde{\alpha}_u^\ast = \frac{23 + 16 n_g}{72 (61 - 16 n_g)} \pi\,,
\end{equation}
which therefore exists for any number of generations $n_g \leq 3$. For $n_g = 3$ we find
\begin{equation}
    \tilde{\alpha}_u^\ast = \frac{71\ \pi}{923}\,.
\end{equation}
While this model 6A is viable in terms of the existence of fixed points, it remains to be studied if their numerical value is compatible with the values of the SM Yukawa couplings, especially for the top Yukawa.

Note that localised Yukawas and generations are also possible in this model, similarly to 5A and with the same features. Furthermore, the $\uu_Z$ symmetry needs to be broken by adding a charged singlet, where the simplest possibility is to include it on the $P_1$ boundary.

\subsection{Model 6A flip on $P_1 \times P_2$}

In this model, $\su 5$ singlets emerge from the fundamentals in the bulk. Furthermore, up and neutrino Yukawas are generated via GHU, hence inheriting the attractive gauge fixed point. In the bulk, we only need two Yukawa couplings:
\begin{equation} \label{eq:Yuk6Aflip}
    \mathcal{L}_\text{Yuk} = - Y_d\ \overline{\Psi}_{20} \Psi_{6} \Phi_{15} - Y_l\ \overline{\Psi}_{\bar 6} \Psi_{20} \Phi_{15} + \text{h.c.}
\end{equation}
This model also conserves baryon number via bulk interaction and features Indalo-particles, where the singlet is part of the bulk ${\bf 6}$.

Labelling $y = {d,\ l}$, the coefficients of the running equations are given by:
\begin{equation}
    c_{yy'} = \begin{pmatrix}
    \frac{29}{6} & 42   \\
    \frac{7}{6} & 174  
    \end{pmatrix}\,, \quad d_y = \begin{pmatrix}
    \frac{287}{12} \\  \frac{287}{12}
    \end{pmatrix}\,.
\end{equation}
For this set-up, the zeros read:
\begin{equation}
    \tilde{\alpha}_d^\ast = \frac{43+64 n_g}{12 (61-16 n_g)} \pi\,, \;\; \tilde{\alpha}_l^\ast = \frac{43+64 n_g}{432 (61-16 n_g)} \pi = \frac{\tilde{\alpha}_d^\ast}{36}\,,
\end{equation}
which are always positive for $n_g \leq 3$. For $n_g$ = 3, we obtain
\begin{equation}
    \tilde{\alpha}^\ast = \frac{6\ \pi}{13}\,, \;\; \tilde{\alpha}_d^\ast = \frac{235\  \pi}{156}\,, \;\; \tilde{\alpha}_l^\ast = \frac{235\ \pi}{5616}\,.
\end{equation}
Regarding localised fields and Yukawas, the same considerations as in 6A apply.

\subsection{Model 6A' on $P_1 \times P_3$}

This model has the same field content as model 6A, except that the different boundary conditions allow only for a up-type Yukawa coupling in the bulk:
\begin{equation} \label{eq:Yuk6Ap}
    \mathcal{L}_\text{Yuk} = - Y_u\ \overline{\Psi}_{\bar{15}} \Psi_{15} \Phi_{15} + \text{h.c.}
\end{equation}
The running is the same as for the model 6A, leading to a fixed point
\begin{equation}
    \tilde{\alpha}_u^\ast = \frac{23 + 16 n_g}{72 (61 - 16 n_g)} \pi\,,
\end{equation}
which therefore exists for any number of generations $n_g \leq 3$.

Besides the presence of a coloured gauge-scalar zero mode, which makes this model non-minimal, the Yukawas for down quarks and leptons can only be added on the boundaries. As an illustration, if we were to add such Yukawas on the $P_1$ boundary, we would be allowed to use $\su 5$ components of the bulk fields whtih have $+$ boundary condition on that boundary (hence, not valishing). Namely:
\begin{eqnarray}
    \Psi_{15} &\rightarrow & \psi_{(10)_1}^1 \supset q_L\,, \;\; \psi^1_{\bar{5}_2} \supset l_L\,; \\
    \Psi_{\bar{15}} &\rightarrow & \psi^2_{(10)_1} \supset u_R^c + e_R^c\,, \;\; \psi^2_{(\bar{5})_{2}} \supset d_R^c\,; \\
    \Phi_{15} &\rightarrow & \phi_{(5)_{-2}} \supset \varphi_H\,;
\end{eqnarray}
where we have used 4D charge conjugation to write all the spinors as left-handed Weyl. The down and  lepton Yukawas can be written as
\begin{equation}
    -\mathcal{L}_\text{Yuk}^\text{loc} = y_d\; \psi^1_{(10)_1} \psi^2_{(\bar{5})_2} \phi_{(5)_{-2}}^\dagger + y_l\; \psi^2_{(10)_1} \psi^1_{(\bar{5})_2} \phi_{(5)_{-2}}^\dagger + \text{h.c.}
\end{equation}
Clearly, these couplings violate $\uu_Z$, having an overall charge of $Q_Z = +5$, hence they are only allowed via the breaking of this symmetry. As a consequence they can only be introduced as higher dimensional operators (dim 5, if we add a localised scalar with charge $q_Z = \pm 5$), or via some heavy mediators. Alternatively, one could add a second Higgs boson, embedded in a localised $\phi_{(5)_{3}}$, hence leading to a type-II 2HDM at low energies, with an additional colour-triplet scalar. In all cases, the introduction of localised Yukawas for down quarks and leptons points toward non-minimal constructions.

\section{Conclusion and perspectives} \label{sec:concl}

Asymptotic Grand Unification offers a novel and intriguing framework for model building, alternative and complementary to traditional quantitative unification. The couplings tend to the same UV fixed point instead of meeting at a fixed energy scale. The presence of UV fixed points for both gauge and Yukawa couplings, combined with the minimality of the low energy model, strongly limits the viable models and allow for a complete and systematic classification of aGUTs.

In this work, we have established a procedure to classify 5D aGUTs, and applied it to models based on a bulk SU(N) gauge symmetry. As a result, we identified two viable and minimal models, whose model building features are summarised in Table~\ref{tab:models2}. Both are based on a bulk $\su 6$, with the same orbifold breaking that allows for a Higgs doublet emerging from the gauge multiplet (GHU). As a consequence, either  the down-type Yukawas ($d$ quarks and charged leptons in 6A) or the up-type Yukawas ($u$ quark and neutrinos in 6A flip) are generated by gauge interactions in the bulk. As such, they run to the same attractive fixed point as the gauge couplings and they are flavour diagonal. Henceforth, flavour mixing can only emerge from the bulk Yukawas with a second Higgs, leading to Minimal Flavour Violation, i.e. one flavour violating matrix in the quark sector and one in the lepton sector, as in the SM. Furthermore, baryon number conservation allows a low compactification scale and the emergence of an asymmetric Dark Matter candidate in the form of the lightest Indalo-particle \cite{Cacciapaglia:2020qky}. A more complete study of the low energy properties of both models is beyond the scope of this work, and it is left for further studies.

\begin{table}[htb]
\centering
\begin{tabular}{|c|c|c|c|c|c|c|c|}
\hline
Name & $\mathcal{G}_\text{bulk}$ & Fermions & Scalars & Yukawas & $n_g$ bulk & Higgs  & UV fixed points ($n_g=3$) \\ \hline \hline
\multicolumn{8}{|c|}{$\mathcal{G}_\text{4D} = \mathcal{G}_\text{SM} \times \uu_Z$} \\ 
\hline
\multirow{2}{*}{6A} & \multirow{2}{*}{$\su 6$} & $\Psi_{15} \supset q_L + d_R\,, \Psi_1 \supset \nu_R\,,$ & $\text{Adj} \supset \varphi_H$ & $d, e$ GHU & \multirow{2}{*}{$3$} & 2HDM & $\tilde{\alpha}^\ast = \tilde{\alpha}_d^\ast = \tilde{\alpha}_l = \frac{6\ \pi}{13}$ \\
 & & $\Psi_{\bar{15}} \supset l_L + u_R + e_R\,, \Psi_{1'}$ & $\Phi_{15} \supset \varphi_H'$ & $u, \nu$ bulk & & Type-II & $\tilde{\alpha}_u^\ast = \frac{415\ \pi}{5538}\,, \;\; \tilde{\alpha}_\nu^\ast = \tilde{\alpha}_\chi^\ast = \frac{122\ \pi}{923}$ \\
\hline
\multirow{2}{*}{6A flip} & \multirow{2}{*}{$\su 6$} & $\Psi_{20} \supset q_L + u_R + e_R\,,$ & $\text{Adj} \supset \varphi_H$ & $u, \nu$ GHU & \multirow{2}{*}{$\leq 3$} & 2HDM & $\tilde{\alpha}^\ast = \tilde{\alpha}_u^\ast = \tilde{\alpha}_\nu^\ast = \frac{6\ \pi}{13}$ \\
 & & $\Psi_{6} \supset d_R\,, \;\; \Psi_{\bar{6}} \supset l_L + \nu_R$ & $\Phi_{15} \supset \varphi_H'$ & $d, e$ bulk & & Type-II & $\tilde{\alpha}_d^\ast = 36\ \tilde{\alpha}_l^\ast = \frac{235\ \pi}{156}$ \\
\hline
\end{tabular}
\caption{\label{tab:models2} Summary of the viable models in the SU(N) kinship with complete UV fixed point for both gauge and Yukawa couplings. Both models enjoy minimal flavour violation, with the mixing matrices stemming from the bulk Yukawas.}
\end{table}

The same procedure can be applied straightforwardly to other classes of bulk gauge symmetries, like SO(N), Sp(N) and exceptional groups. However, viable models can also be obtained by releasing some of the minimality requirements. \emph{In primis}, the requirement that the 4D remnant should correspond to the SM gauge symmetry could be lifted in favour of other realistic and motivated cases. The only requirement is that unification does not occur, in the sense that one can always match the SM gauge couplings to those of the intermediate step. Pati-Salam \cite{Pati:1974yy}, based on $\su{4} \times \su{2}_\text{L} \times \su{2}_\text{R}$, is an attractive case as it requires a minimal fermion content matching that of the SM. An aGUT with intermediate Pati-Salam was proposed in Ref.~\cite{Cacciapaglia:2023ghp}. Another possibility consists of Left-Right symmetric models \cite{Mohapatra:1974gc,Mohapatra:1974hk,Senjanovic:1975rk}, based on $\su{3}_c \times \su{2}_\text{L} \times \su{2}_\text{R} \times \uu_\text{B-L}$, originally proposed to restore parity in the SM and to generate neutrino masses effectively \cite{Davidson:1978pm,Mohapatra:1980qe}. Other possibilities include 331 models \cite{Frampton:1992wt,Pisano:1992bxx}, based on $\su{3}_c \times \su{3}_\text{L} \times \uu$, and trinification \cite{Babu:1985gi,CarcamoHernandez:2020owa}, based on $\su{3}_c\times \su{3}_\text{L}\times \su{3}_\text{R}$, which could explain the presence of three generations via gauge anomaly cancellation. Another intriguing possibility would be the inclusion of gauged flavour symmetries, which have been recently reconsidered as a consistent scenario of new physics \cite{Grinstein:2010ve,DAgnolo:2012ulg,Darme:2023nsy}. The presence of extended gauge symmetries in 4D, however, poses the additional challenge of breaking them down to the SM, hence requiring additional fields to be added in the bulk or on the boundary.

This first systematic analysis of SU(N) aGUTs in 5D illustrates the great potential of this class of models. An attractive feature is due to the strong requirements from the UV fixed points, which greatly limits the model building possibilities, while still allowing for interesting phenomenological consequences and predictions.

\section*{Acknowledgements}

The author is indebted to Corentin Cot for his PhD thesis notes, which were crucial for the computation of the Yukawa coupling running.

\appendix

\section{Comment on modified  boundary conditions} \label{app:BCs}

The parities under the two $\mathbb{Z}_2$ symmetries defining the orbifold can be translated into boundary conditions at the edges of the interval defining the extra dimensions. As we will see, when such conditions stem from an orbifold, all bulk couplings generated by the bulk gauge symmetry are consistently preserved. It has been proposed that more general boundary conditions could be applied, however, which can further reduce the 4D remnant gauge invariance or change the zero mode patterns \cite{Csaki:2003dt,Csaki:2003sh}. This technique has been used to define models in 5D, see for instance Ref.~\cite{Angelescu:2021nbp}. We will see with a simple example that the latter is not compatible with the full 5D gauge invariance, as some couplings in the bulk are forced to vanish, while they are required by the bulk gauge group.

As a concrete example, let's consider an $\su 5$ model based on the parities $P_1 \times P_1'$. As discussed in Sec.~\ref{subsec:su5}, this orbifold always leads to unwanted chiral zero modes. For instance, a ${\bf 5}$ in the bulk contains the following SM components:
\begin{equation}
    {\bf 5} \to (3,1)_{-1/3} \oplus (1,2)_{1/2}\,.
\end{equation}
Assigning parities $(-,-)$, the bulk field will contain a right-handed zero mode in the first, matching a right-handed down singlet, but also a left-handed zero mode in the second. The latter has conjugate quantum numbers with respect to the SM lepton doublet. The bulk gauge interactions of the ${\bf 5}$ fermion read, in SM components:
\begin{equation}
    \overline{\Psi}_5 \gamma_\mu D^\mu_5 \Psi_5 = \overline{\Psi}_d \gamma_\mu D^\mu \Psi_d + \overline{\Psi}_{l^c} \gamma_\mu D^\mu \Psi_{l^c} - i g_5\ \left(\overline{\Psi}_d \gamma_\mu A^\mu_{(3,2)_{-5/6}} \Psi_{L^c} + \text{h.c.} \right)\,,
\end{equation}
where the covariant derivative on the right-hand-side contains the SM gauge bosons. The parities of the components are consistent with the presence of all such terms, including the couplings of the gauge field $A^\mu_{(3,2),_{-5/6}}$.

To remove the zero mode in the $\Psi_{l^c}$ component, one can add a localised right-handed spinor with the same quantum numbers, $\psi_{rh,l^c}$, and couple the two via a mass term
\begin{equation}
    \mathcal{L}_\text{mass} = M_\text{loc}\ \delta (x_5) \ \overline{\Psi}_{l^c} \psi_{rh,l^c} + \text{h.c.}
\end{equation}
In the limit of large $M_\text{loc} \to \infty$, the mass term effectively forces the left-handed component of $\Psi_{l^c}$ to vanish, hence flipping the boundary condition on the field from $(-,-) \to (+,-)$. Now we can see that the couplings of $A^\mu_{(3,2)_{-5/6}}$ become odd under the orbifold, hence all the KK couplings stemming from such term vanish. This effect, therefore, violates explicitly the bulk $\su 5$ symmetry. While such coupling does not appear at one loop in the running of the gauge couplings, it will affect the one loop running of the Yukawa couplings and also the running at higher orders. Hence, the $\su 5$--invariant behaviour at high energies, including the fixed point, will be destabilised. 

For the reasons illustrated above, we cannot consider such scenario in the aGUT model building. One possible variant consists in keeping the localised mass term finite. In this way, at high scales $\mu \gg M_\text{loc}$, a consistent orbifold is recovered. Henceforth, one could also consider models that do not lead to a consistent low energy theory matching the SM and only modify the low energy spectrum below $M_\text{loc}$. The price to pay, however, is a non-minimal set up for the model, hence we will not consider this option in this work.

\section{Evolution of the Yukawa couplings}\label{app:Yukawas}

The results shown here are mainly adapted from Ref.~\cite{Cot2021}. We provide details on how the coefficient $c_{yy'}$ and $d_y$ are extracted. 

In general, a Yukawa coupling will have the following structure:
\begin{equation}
    \mathcal{L}_\text{Yuk} \supset Y_k\ \overline{\Psi}_B \Psi_A \Phi_C + \text{h.c.}
\end{equation}
where $A,B,C$ label irreps of SU(N). Hence, it is convenient to write the Yukawa coupling as a matrix carrying the gauge indices of such irreps, $\alpha, \beta, \gamma$:
\begin{equation}
    Y_k \equiv y_k\ Y^{\alpha, \beta, \gamma}\,.
\end{equation}
In the following, we will threat the $Y$'s as matrices in the gauge space. As a general notation, greek letters indicate, generically, indices of different irreps, while roman letters will indicate the gauge indices of the fundamental irrep. Furthermore, roman superscripts indicate indices of the fundamental irrep, and subscripts of the anti-fundamental.

In the models of concern, there are several templates, with the following structures:
\begin{eqnarray}
    \overline{\Psi}_F \Psi_0 \Phi_F\; (\overline{\Psi}_0 \Psi_{\bar{F}} \Phi_F) &\rightarrow & Y^{\alpha, \beta, \gamma} = {Y_{a}}^{b} = \delta^b_a\,,\\
    \overline{\Psi}_A \Psi_0 \Phi_A\; (\overline{\Psi}_0 \Psi_{\bar{A}} \Phi_A) &\rightarrow & Y^{\alpha, \beta, \gamma} = {Y_{a_1 a_2}}^{b_1 b_2} = \frac{1}{2} \left(\delta^{b_1}_{a_1} \delta^{b_2}_{a_2} - \delta^{b_1}_{a_2} \delta^{b_2}_{a_1} \right)\,, \\
    \overline{\Psi}_A \Psi_F \Phi_F\; (\overline{\Psi}_{\bar{F}} \Psi_{\bar{A}} \Phi_F) &\rightarrow & Y^{\alpha, \beta, \gamma} = {{Y^{a}}_{b_1 b_2}}^{c} = \frac{1}{2} \left(\delta^{a}_{b_1} \delta^{c}_{b_2} - \delta^{a}_{b_2} \delta^{c}_{b_1} \right)\,, \\
    \overline{\Psi}_{A_3} \Psi_F \Phi_A\; (\overline{\Psi}_{\bar{F}} \Psi_{\bar{A}_3} \Phi_A) &\rightarrow & Y^{\alpha, \beta, \gamma} = {{Y^{a}}_{b_1 b_2 b_3}}^{c_1 c_2} = \frac{1}{6} \left(\delta^{a}_{b_1} \delta^{c_1}_{b_2} \delta^{c_2}_{b_3} - \text{perm.} (b_1,b_2,b_3) \right)\,.
\end{eqnarray}
There are also three special cases, which only apply to $\su 5$ or $\su 6$ models:
\begin{eqnarray}
    \text{SU(5) :}\;\; \overline{\Psi}_{\bar{A}} \Psi_A \Phi_F \rightarrow Y^{\alpha,\beta,\gamma} = Y^{a_1 a_2, b_1 b_2, c} = \epsilon_5^{a_1 a_2 b_1 b_2 c}\,, \\
    \text{SU(6) :}\;\; \overline{\Psi}_{\bar{A}} \Psi_A \Phi_A \rightarrow Y^{\alpha,\beta,\gamma} = Y^{a_1 a_2, b_1 b_2, c_1 c_2} = \epsilon_6^{a_1 a_2 b_1 b_2 c_1 c_2}\,, \\
    \text{SU(6) :}\;\; \overline{\Psi}_{\bar{F}} \Psi_{A_3} \Phi_A \rightarrow Y^{\alpha,\beta,\gamma} = Y^{a, b_1 b_2 b_3, c_1 c_2} = \epsilon_6^{a b_1 b_2 b_3 c_1 c_2}\,.
\end{eqnarray}

\subsection{Gauge contribution}

In the 5D model, both the 4D vector and scalar components contribute. The contributions are computed in a general $R_\xi$ gauge, where $\xi$-dependent terms should cancel out to obtain the well-known one-loop gauge-independent results.

From the results in Ref.~\cite{Cot2021}, we extracted the following master formula:
\begin{multline} \label{eq:MasterGauge}
    d_y Y^{\alpha,\beta,\gamma} = (8 + 2 \xi)\ T^\alpha_{\alpha'} T^\beta_{\beta'} Y^{\alpha',\beta',\gamma} - (2 \xi)\ T^\alpha_{\alpha'} T^\gamma_{\gamma'} Y^{\alpha',\beta,\gamma'} + (2 \xi)\ T^\alpha_{\alpha'} T^\gamma_{\gamma'} Y^{\alpha',\beta,\gamma'} + \\
    - \left(\xi+\frac{1}{2} \right) \left( C_2(R_A) + C_2(R_B) \right)\ Y^{\alpha,\beta,\gamma} - (\xi-3) C_2(R_C)\ Y^{\alpha,\beta,\gamma}\,.
\end{multline}
In the formula above, $C_2(R_X)$ is the Casimir of the representation $X$, and $T$ indicate the gauge generators in the irrep carrying the proper index (a sum is left intended for repeated generators). Note that $T \to -T$ for the conjugate irreps.

We express all generators in terms of fundamental indices:
\begin{eqnarray}
    T_F &=& T^i_j\,, \\
    T_A &=& T^{ij}_{lk} = T^{i}_{p} \left(\delta^p_l \delta^j_k - \delta^p_k \delta^j_l  \right)\,, \\
    T_{A_3} &=& T^{ijy}_{lkm} = T^{i}_{p} \frac{1}{2} \left(\delta^p_l \delta^j_k \delta^y_m - \text{perm.} (l,k,m) \right)\,. 
\end{eqnarray}
We will also use the relation:
\begin{equation}
    \sum_T T^i_j T^l_k = \frac{1}{2} \delta^i_k \delta^l_j - \frac{1}{2N} \delta^i_j \delta^l_k\,,
\end{equation}
which leads to
\begin{equation}
    \sum_T T^i_{i'} T^j_{j'} \left(\delta^{i'}_l \delta^{j'}_k - \delta^{i'}_k \delta^{j'}_l \right) = - \frac{N+1}{N}\ \left(\delta^i_l \delta^j_k - \delta^i_k \delta^j_l \right)\,, 
\end{equation}
and the identities:
\begin{equation}
    -\frac{N+1}{N} = C_2(A) - 2 C_2(F) = \frac{1}{2} (C_2(A_3) - C_2(A) - C_2(F))\,,
\end{equation}
and
\begin{equation}
    C_2(A_3) = 3 (C_2(A) - C_2(F))\,.
\end{equation}

\subsubsection{Template $\overline{\Psi}_R \Psi_0 \Phi_R$}

In this case:
\begin{equation}
    T^\beta_{\beta'} T^\gamma_{\gamma'} Y^{\alpha,\beta',\gamma'} = C_2 (R)\ Y^{\alpha,\beta,\gamma}\,.
\end{equation}
Hence, from Eq.~\eqref{eq:MasterGauge}, we obtain:
\begin{equation}
    d_y = \frac{5}{2} C_2(R)\,,
\end{equation}
where the $\xi$-dependent terms cancel out. The same result is obtained for the companion template $\overline{\Psi}_0 \Psi_{\bar{R}} \Phi_R$.

For the models of concern, we have:
\begin{eqnarray}
    \text{Model 5A},\; R=F \rightarrow d_y = 6\,, \\
    \text{Model 6A},\; R=A \rightarrow d_y = \frac{35}{3}\,. 
\end{eqnarray}

\subsubsection{Template $\overline{\Psi}_A \Psi_F \Phi_F$}

In this case:
\begin{eqnarray}
    T^\alpha_{\alpha'} T^\beta_{\beta'} Y^{\alpha',\beta',\gamma} &=&  \frac{1}{2}\ C_2(A)\ Y^{\alpha,\beta,\gamma}\,, \\
    T^\alpha_{\alpha'} T^\gamma_{\gamma'} Y^{\alpha',\beta,\gamma'} &=& - \frac{N+1}{N} \frac{1}{2}\ Y^{\alpha,\beta,\gamma}\,, \\
    T^\beta_{\beta'} T^\gamma_{\gamma'} Y^{\alpha,\beta',\gamma'} &=&  \frac{1}{2}\ C_2(A)\ Y^{\alpha,\beta,\gamma}\,.
\end{eqnarray}
From Eq.~\eqref{eq:MasterGauge}, we obtain
\begin{equation}
    d_y = \frac{7 C_2(A) + 5 C_2(F)}{2}\,.
\end{equation}
For the models of concern, we have:
\begin{eqnarray}
    \text{Model 5A} \rightarrow d_y = \frac{93}{5}\,.
\end{eqnarray}

\subsubsection{Template $\overline{\Psi}_{A_3} \Psi_F \Phi_A$}

In this case:
\begin{eqnarray}
    T^\alpha_{\alpha'} T^\beta_{\beta'} Y^{\alpha',\beta',\gamma} &=&  \frac{1}{3}\ C_2(A_3)\ Y^{\alpha,\beta,\gamma}\,, \\
    T^\alpha_{\alpha'} T^\gamma_{\gamma'} Y^{\alpha',\beta,\gamma'} &=& \frac{2}{3}\ C_2 (A_3)\ Y^{\alpha,\beta,\gamma}\,, \\
    T^\beta_{\beta'} T^\gamma_{\gamma'} Y^{\alpha,\beta',\gamma'} &=&  -\frac{N+1}{N}\ Y^{\alpha,\beta,\gamma}\,.
\end{eqnarray}
From Eq.~\eqref{eq:MasterGauge}, we obtain
\begin{equation}
    d_y = \frac{19 C_2(A) - 14 C_2(F)}{2}\,.
\end{equation}
For the models of concern, we have:
\begin{eqnarray}
    \text{Model 6A flip} \rightarrow d_y = \frac{287}{12}\,.
\end{eqnarray}

\subsubsection{Special $\su 5$ template $\overline{\Psi}_{\bar{A}} \Psi_A \Phi_F$}

In this case (we include explicitly the minus signs from the $\bar{A}$, hence $T^\beta_{\beta'} \to - T^\beta_{\beta'}$):
\begin{eqnarray}
    -T^\alpha_{\alpha'} T^\beta_{\beta'} Y^{\alpha',\beta',\gamma} &=&  2 \frac{N+1}{N}\ Y^{\alpha,\beta,\gamma}\,, \\
    T^\alpha_{\alpha'} T^\gamma_{\gamma'} Y^{\alpha',\beta,\gamma'} &=& - \frac{N+1}{N}\ Y^{\alpha,\beta,\gamma}\,, \\
    -T^\beta_{\beta'} T^\gamma_{\gamma'} Y^{\alpha,\beta',\gamma'} &=&  \frac{N+1}{N}\ Y^{\alpha,\beta,\gamma}\,.
\end{eqnarray}
From Eq.~\eqref{eq:MasterGauge}, we obtain
\begin{equation}
    d_y = -17 C_2(A) + 35 C_2(F) - \xi \left( 10 C_2(A)-15 C_S(F)\right)\,.
\end{equation}
The  $\xi$-dependent term only cancels for $N=5$, and for the models of concern, we have:
\begin{eqnarray}
    \text{Model 5A} \rightarrow d_y = \frac{114}{5}\,.
\end{eqnarray}

\subsubsection{Special $\su 6$ template $\overline{\Psi}_{\bar{A}} \Psi_A \Phi_A$}

In this case (we include explicitly the minus signs from the $\bar{A}$, hence $T^\beta_{\beta'} \to - T^\beta_{\beta'}$):
\begin{eqnarray}
    -T^\alpha_{\alpha'} T^\beta_{\beta'} Y^{\alpha',\beta',\gamma} &=&  2 \frac{N+1}{N}\ Y^{\alpha,\beta,\gamma}\,, \\
    T^\alpha_{\alpha'} T^\gamma_{\gamma'} Y^{\alpha',\beta,\gamma'} &=& - 2 \frac{N+1}{N}\ Y^{\alpha,\beta,\gamma}\,, \\
    -T^\beta_{\beta'} T^\gamma_{\gamma'} Y^{\alpha,\beta',\gamma'} &=&  2 \frac{N+1}{N}\ Y^{\alpha,\beta,\gamma}\,.
\end{eqnarray}
From Eq.~\eqref{eq:MasterGauge}, we obtain
\begin{equation}
    d_y = -14 C_2(A) + 32 C_2(F) - \xi \left( 15 C_2(A)-24 C_S(F)\right)\,.
\end{equation}
The  $\xi$-dependent term only cancels for $N=5$, and for the models of concern, we have:
\begin{eqnarray}
    \text{Model 6A} \rightarrow d_y = 28\,.
\end{eqnarray}

\subsubsection{Special $\su 6$ template $\overline{\Psi}_{\bar{F}} \Psi_{A_3} \Phi_A$}

In this case (we include explicitly the minus signs from the $\bar{A}$, hence $T^\beta_{\beta'} \to - T^\beta_{\beta'}$):
\begin{eqnarray}
    -T^\alpha_{\alpha'} T^\beta_{\beta'} Y^{\alpha',\beta',\gamma} &=&  \frac{3}{2} \frac{N+1}{N}\ Y^{\alpha,\beta,\gamma}\,, \\
    T^\alpha_{\alpha'} T^\gamma_{\gamma'} Y^{\alpha',\beta,\gamma'} &=& - 3 \frac{N+1}{N}\ Y^{\alpha,\beta,\gamma}\,, \\
    -T^\beta_{\beta'} T^\gamma_{\gamma'} Y^{\alpha,\beta',\gamma'} &=&  \frac{N+1}{N}\ Y^{\alpha,\beta,\gamma}\,.
\end{eqnarray}
From Eq.~\eqref{eq:MasterGauge}, we obtain
\begin{equation}
    d_y = -\frac{21}{2} C_2(A) + 25 C_2(F) - \xi \left( 15 C_2(A)-24 C_S(F)\right)\,.
\end{equation}
The  $\xi$-dependent term only cancels for $N=5$, and for the models of concern, we have:
\begin{eqnarray}
    \text{Model 6A} \rightarrow d_y = \frac{287}{12}\,.
\end{eqnarray}

\subsection{Yukawa contributions}

These effects can be divided into two types: loops on the external legs, and triangle loops.

For the former, all Yukawas that contain the given field contribute. For a Yukawa of the form $\overline{\Psi}_A \Psi_B \Phi_C$, the effect can be expressed as:
\begin{equation}
    c_{yy'}^{(1)} = \frac{1}{2} \zeta_{y'}^A - \frac{1}{2} \zeta_{y'}^B + 2 \zeta_{y'}^C\,,
\end{equation}
where $\zeta_{y'}^X$ is the correction of the Yukawa $y'$ on the field line $X$. We can compute the coefficients $\zeta$ for all Yukawa templates:
\begin{eqnarray}
    \overline{\Psi}_R \Psi_0 \Phi_R &\rightarrow & \zeta^0 = d(R)\,, \;\; \zeta^R = 1\,; \\
    \overline{\Psi}_A \Psi_F \Phi_F &\rightarrow & \zeta^F = \frac{1}{2} (d(F)-1)\,, \;\; \zeta^A = 1\,; \\
    \overline{\Psi}_{A_3} \Psi_F \Phi_A &\rightarrow & \zeta^F = \frac{1}{3} (d(A) - d(F) +1)\,, \;\; \zeta^A = \frac{1}{3} (d(F)-2)\,, \;\; \zeta^{A_3} = 1\,; \\
    \overline{\Psi}_{\bar{A}} \Psi_A \Phi_F &\rightarrow & \zeta^A = 12\,, \;\; \zeta^F = 24\,; \\
    \overline{\Psi}_{\bar{A}} \Psi_A \Phi_A &\rightarrow & \zeta^A = 48\,; \\
    \overline{\Psi}_{\bar{F}} \Psi_{A_3} \Phi_A &\rightarrow & \zeta^{A_3} = 36\,, \;\; \zeta^A = 48\,, \;\; \zeta^F = 120\,,
\end{eqnarray}
where $d(F) = N$ and $d(A) = (N^2-N)/2$ are the dimensions of the irreps, and the same coefficient applies to the conjugate irreps.

The second contribution stems from triangle loops. In general, the effect stems from two possible topologies, leading to the master formula:
\begin{equation}
    c_{yy'}^{(2)} Y^{\alpha,\beta,\gamma} = (-2) \times \left[ (Y')^\dagger_{\alpha,\delta,\gamma'} (Y')^{\alpha',\delta,\gamma} Y^{\alpha',\beta,\gamma'} \;\; \text{or} \;\; Y^{\alpha,\beta',\gamma'} (Y')^{\beta',\delta,\gamma} (Y')^\dagger_{\beta,\delta',\gamma'} \right]\,,
\end{equation}
where sums over the repeated indices indicate a trace over the indices of the corresponding irrep.
In the models of interest, the traces usually involve two different Yukawas, and must be evaluated case by case.

\bibliographystyle{jhep}
\bibliography{biblio}

\end{document}